\documentclass[conference]{IEEEtran}
\IEEEoverridecommandlockouts

\usepackage[font=footnotesize]{caption}
\usepackage{cite}
\usepackage{amsmath,amssymb,amsfonts}
\usepackage{algorithmic}
\usepackage{graphicx}
\usepackage{textcomp}
\usepackage{xcolor}
\usepackage{booktabs}
\usepackage[ruled,vlined]{algorithm2e}
\usepackage{enumitem}
\usepackage[hyphens]{url}
\usepackage{wrapfig}
\usepackage[hidelinks]{hyperref}
\usepackage{subcaption}
\usepackage{adjustbox}
\usepackage{etoolbox}

\def\BibTeX{{\rm B\kern-.05em{\sc i\kern-.025em b}\kern-.08em
    T\kern-.1667em\lower.7ex\hbox{E}\kern-.125emX}}
\begin{document}

\pdfpagewidth=8.5in
\pdfpageheight=11in

\newcommand{\sto}[1][3pt]{\mathrel{%
  \hbox{\rule[\dimexpr\fontdimen22\textfont2-.2pt\relax]{#1}{.4pt}}%
  \mkern-4mu\hbox{\usefont{U}{lasy}{m}{n}\symbol{41}}}}

\newcommand{\alice}[1]{}
\newcommand{\hanyu}[1]{}
\newcommand{\revtag}[2]{}
\newcommand{\revnote}[2]{}
\newcommand{\revcam}[2][]{#2}
\newcommand{\reva}[2][]{#2}
\newcommand{\revb}[2][]{#2}
\newcommand{\revc}[2][]{#2}
\newcommand{\revd}[2][]{#2}
\newcommand{\reve}[2][]{#2}
\newcommand{\revf}[2][]{#2}
\newcommand{\revall}[2][]{#2}

\title{SegFold: Accelerating Sparse GEMM with a Fine-Grained Dynamic Dataflow}
\author{
\IEEEauthorblockN{Xinrui Wu, Hanyu Wang, Jason Cong, and Tony Nowatzki}
\IEEEauthorblockA{
University of California, Los Angeles, USA\\
\{alicewu,hanyuwang,cong,tjn\}@cs.ucla.edu
}
}

\maketitle

\begin{abstract}
Generalized sparse matrix-matrix multiplication (SpGEMM) is critical in many domains.
Existing CPUs and GPUs, as well as specialized accelerators, rely on static dataflows (e.g., inner product, outer product, Gustavson, etc.).  Each static dataflow sacrifices some data reuse opportunities and imposes constraints on load balance.

To address this inefficiency, we extend the typical SpGEMM dataflows by considering dynamism. Specifically, we add fine-grained dynamic scheduling to optimize reuse and reduce resource contention. We also develop dynamic remapping of partially completed work to improve load balance and parallelism.
These ideas are formalized into a specific dataflow called Segment.
To demonstrate Segment, we codesign a SpGEMM accelerator called SegFold.
SegFold includes a memory controller that identifies fine-grained reuse opportunities in a local window of the stationary input array and exploits them through dynamic work assignment. It also incorporates a merge network that routes inputs to appropriate processing elements (PEs) for computation while dynamically remapping the work assigned to each PE to balance load.
Across diverse densities and matrix sizes, SegFold achieves a geometric-mean $1.95\times$ speedup over state-of-the-art SpGEMM accelerators and $5.3\times$ over the best static dataflow configuration, demonstrating that adding dynamism to the dataflow design space unlocks reuse and load-balance gains that no static scheduling choice can achieve in isolation.
\end{abstract}


\section{Introduction}
Generalized sparse matrix-matrix multiplication (SpGEMM) is rapidly becoming the algorithmic backbone of workloads with dual-side sparsity, where both operands are sparse. Such dual-side sparsity arises not only in traditional domains—e.g., graph analytics, scientific simulation, optimization, and economic modeling—but also in modern machine learning, including pruned or structurally sparse DNN inference and sparse attention in large language models.

SpGEMM is a difficult workload to execute efficiently, as the computations are fundamentally data-dependent at a fine-grain, resulting in a host of challenges from irregular memory access, data-dependent control, load imbalance, and excessive bandwidth use.  These challenges are particularly severe for unstructured sparsity, which is our focus.
The vector execution models on CPUs and GPUs are not well-suited for these data-dependent computations, and thus a wide design space of specialized sparse matrix accelerators has been explored (e.g.~\cite{svm-accel, outerspace, cambricon-x, cambricon-s, scnn, extensor, matraptor,sparten,tensaurus,feasta,multi-compression,koul2024onyx,song2022sextans,he2020sparse,song2022serpens}).  However, prior accelerators still have large headroom for improvement, as they either require high hardware overhead or do not achieve high utilization. 

While the SpGEMM design space has been well formalized and explored (e.g.~\cite{teaal,sparseloop,flexagon,stellar, daveSurvey,spgemm-survey}), the conventional taxonomy ignores a key dimension of dynamism, leading to lost opportunities in achieving high data reuse and good load balance across processing elements (PEs)---both of which are key to high efficiency.

Prior accelerators typically adopt one of three dataflows to process each tile: inner product, outer product, or Gustavson. Recent designs such as Spada~\cite{spada}, Trapezoid~\cite{trapezoid}, and Flexagon~\cite{flexagon} can select among these per tile, but the scheduling within each tile remains static. Additionally, each of the three dataflows makes different tradeoffs. Inner product reuses partial sums but sacrifices input locality and requires balanced row-column intersections. Outer product reuses inputs but sacrifices output reuse and requires balanced input nonzeros. Gustavson sacrifices reuse on one of the two inputs and imposes some load balance constraints on both nonzeros and intersections. The best approach depends on the sparsity pattern of the dataset~\cite{flexagon,sparGD,dynaflow}, and even depends on sparsity within active tiles~\cite{spada}.

Our key insight is that higher reuse and better load balance can be achieved by introducing temporal and spatial dynamism into the dataflow for processing a single tile.  First, instead of choosing a static ordering of rows and columns to process, a dataflow with \emph{dynamic scheduling} can choose different fine-grained execution orders on different tiles to improve reuse and reduce resource contention.  A dataflow with \emph{dynamic mapping} can also re-partition storage dynamically to optimize for high PE utilization.  These degrees of dynamism show that the current dataflow taxonomies are incomplete.

In this work, we propose a new dynamic dataflow and codesign an efficient accelerator. Our dataflow, called Segment, uses dynamic scheduling to select a set of elements and rows from input matrices to multiply in each cycle; this dynamic ordering enables Segment to achieve the benefits of both \textit{outer product} and \textit{Gustavson} dataflows.  Segment further balances load through dynamic mapping of partial sums across PEs; PEs redistribute partial sums across the array based on intersection patterns while preserving column ordering.  Both aspects enable high utilization.

To implement this novel dataflow, we develop our microarchitecture named SegFold. SegFold comprises three key strategies: the memory controller (for dynamic scheduling), the adaptive merge network (for dynamic mapping), and the folding mechanism (for virtualizing hardware resources).
For dynamic scheduling, the memory controller maintains an active window to monitor multiple input rows and assign resources to maximize reuse within the window on-the-fly.
The merge network enables dynamic mapping by redistributing each computation to an available PE based on the current distribution of partial sums. 
Finally, output rows vary in length: long rows can overflow a single physical PE row, while short rows leave neighboring PE rows underutilized. We implement a folding mechanism that maps overflow portions of long output rows onto neighboring physical PE rows whose resident output rows are shorter, improving utilization without requiring each logical row to fit within one PE row.

We created a detailed cycle-level simulator for SegFold, and implemented its major components in RTL for resource evaluation. We evaluate SegFold on a diverse set of matrices with different sizes and sparsity levels. 
We compare SegFold against state-of-the-art SpGEMM accelerators that support multiple static dataflows~\cite{flexagon} or adaptive static dataflows~\cite{spada}. Overall, SegFold achieves a geometric-mean speedup of $1.95\times$ over Spada and $5.3\times$ over the best Flexagon dataflow configuration.

To summarize, our main contributions are:
\begin{itemize}
    \item A novel dataflow aspect: dynamic scheduling that reorders work in an active window to maximize data reuse.
    \item A novel dataflow aspect: dynamic mapping that distributes partial sums across PEs to balance load while preserving column ordering.
    \item A codesigned efficient microarchitecture with significant speedups over conventional static dataflows across a range of input matrices and sparsity levels.
\end{itemize}

\section{Background}\label{sec:background}
\subsection{SpGEMM Dataflow Design Space}
We use $A, B$ to denote the input of SpGEMM, and $C$ for the output. The arithmetic representation for SpGEMM is:
\begin{align}
    \label{eq:spgemm}
    C_{m,n} = \textstyle \sum_{k}A_{m,k}\times B_{k,n},
\end{align}

where both $A$ and $B$ are sparse.
The iteration space
\begin{align*}
    S = \{(m,n,k)\ |\ 0\leq m< M,\ 0\leq n < N,\ 0\leq k < K\},
\end{align*}
can be broken into multiplications and additions as:
\begin{align}
    \label{eq:mul}
    T_{m,n,k} = A_{m,k}\times B_{k,n}, \\
    \label{eq:add}
    C_{m,n} = \textstyle \sum_{k}T_{m,n,k}.
\end{align}

$T$ is the product of $A$ and $B$ elements paired by the shared index $k$; $C$ is the reduction of $T$ along the $K$ dimension.

The major aspects of SpGEMM's dataflow design are:
\begin{enumerate}
    \item Tiling strategy for $M$, $N$ and $K$.
    \item Scheduling of all dimensions ($M$, $N$, $K$ plus tiling subdimensions) onto spatial and temporal axes.
    \item Fusion of the computation in Eq.~\ref{eq:mul} and Eq.~\ref{eq:add}.
\end{enumerate}

Our work focuses on the second and third aspects -- i.e., the scheduling and fusion of work at a fine-grain within a tile. In this paper, we discuss three representative dataflows—inner product, outer product, and Gustavson. We next explain how each dataflow uses different scheduling and fusion strategies.

\begin{figure*}[t]
    \centering
    \includegraphics[width=0.98\linewidth]{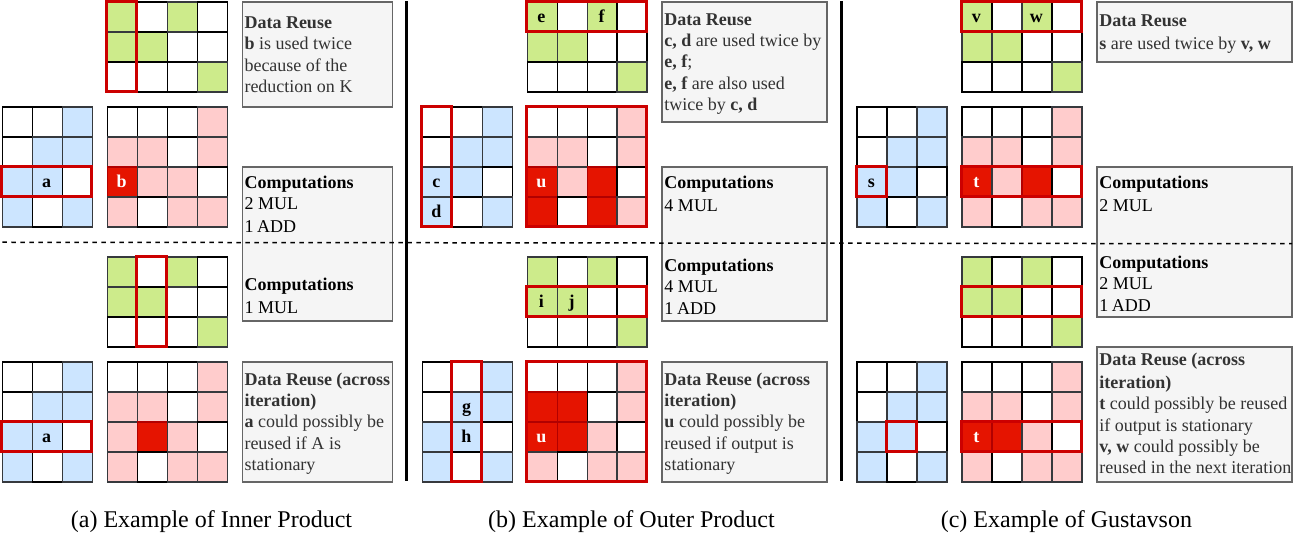}
    \vspace{-0.1in}
    \caption{Comparison of classic dataflows. Every two consecutive blocks  in a column (top vs bottom) are two sequential snapshots. }
    \label{fig:dataflow-bg}
    \vspace{-0.09in}
\end{figure*}

\smallskip\noindent\textbf{Static Dataflow Definitions.}
\emph{Inner product} (order $M \!\sto\! N \!\sto\! K$) performs dot products between rows of $A$ and columns of $B$. 
\emph{Outer product} (order $K \!\sto\! M \!\sto\! N$) performs cross products between column vectors from $A$ and row vectors from $B$.
\emph{Gustavson} (order $M \!\sto\! K\!\sto\! N$) performs row products between elements $A_{m,k}$ and the corresponding $k$ row from $B$. Figure~\ref{fig:dataflow-bg} shows an example of these dataflows, which we discuss further in $\S$\ref{sec:static-limitation} to elucidate their limitations.

\subsection{Dataflow in Prior Work}

\begin{table}[!t]
\footnotesize
\centering
\caption{Taxonomy of SpGEMM accelerators.
}
\begin{tabular}{@{}lp{2cm}p{4cm}@{}}
\toprule
\textbf{Accelerator} & \textbf{Scheduling} & \textbf{Reconfigurability to\newline Balance Load \& Comp.} \\
\midrule
TPU~\cite{tpu}               & -- & -- \\
ExTensor~\cite{extensor}     & IP & -- \\
SIGMA~\cite{sigma}           & IP & Reconfigurable reduction network \\
OuterSpace~\cite{outerspace} & OP & None \\
SpArch~\cite{sparch}         & OP & Data-dependent comparator arrays \\
MatRaptor~\cite{matraptor}   & Gust & Comparator queues \\
Gamma~\cite{gamma}           & Gust & None \\
Flexagon~\cite{flexagon}     & IP/OP/Gust & Reconfigurable distribution, merge, and reduction networks \\
Trapezoid~\cite{trapezoid}   & IP/Gust (spatial/temporal) & Reconfigurable distribution, merge, and reduction networks \\
Spada~\cite{spada}           & Window/Tile-size Adaptive & Neighboring-lane work stealing\\

\textbf{SegFold (ours)} & Sub-tile Dynamic dataflow & Sub-tile scheduling \& Reconfigurable merge network \\
\bottomrule
\end{tabular}
\vspace{-0.24in}
\label{tab:taxonomy}
\end{table}

SpGEMM accelerators span a diverse design space, characterized by their dataflow strategies and degrees of hardware reconfigurability (see Table~\ref{tab:taxonomy}). Early versions of TPU~\cite{tpu,tpu2} use weight-stationary systolic arrays optimized for dense workloads, without sparsity support.

In the sparse domain, most prior accelerators adopt a static dataflow. Architectures like ExTensor adopt a fixed inner-product dataflow~\cite{extensor}. SIGMA builds upon this by incorporating a reconfigurable reduction network~\cite{sigma}. Outer-product accelerators, such as OuterSpace~\cite{outerspace} and SpArch~\cite{sparch}, exploit sparsity using mechanisms like specialized memory hierarchy or comparator arrays. Gustavson-inspired designs, including MatRaptor~\cite{matraptor} and Gamma~\cite{gamma}, focus on row-/column-wise intersections through comparator queues; Gamma and Zed~\cite{zed} additionally include a matrix-preprocessing step that groups similar rows of the stationary matrix to improve reuse on the streaming matrix.

Recent efforts like
Flexagon~\cite{flexagon},
Trapezoid~\cite{trapezoid},
SpMARD~\cite{spmard},
SparGD~\cite{sparGD}, and
SPARM~\cite{sparm}
support multiple dataflows (inner-product, outer-product, Gustavson), employing reconfigurable distribution, merge, and reduction fabrics for greater flexibility across different matrices.

Spada~\cite{spada} develops a window-adaptive (WA) dataflow that supports a spectrum of execution modes by adjusting window height and width at tile granularity to realize the input and output reuse benefits of different dataflows under varying sparse patterns. It uses local dynamism: neighboring lanes can opportunistically process each other's elements when idle, providing some degree of dynamic load balancing within the merge network. However, the overall scheduling remains statically determined by the tiled loop structure.

Within this continuum, SegFold introduces a dynamic dataflow that adapts at sub-tile granularity. Unlike Spada's window-size adaptation, SegFold dynamically reorders work selection (\textsc{SelectA} in $\S$\ref{sec:selectA}) within an active window based on instantaneous reuse opportunities, and uses a reconfigurable merge network to dynamically remap partial sums (\textsc{SegmentBC} in $\S$\ref{sec:segmentBC}) based on output sparsity patterns discovered on-the-fly. 

\subsection{Limitations of Static Scheduling: Low Data Reuse and Imbalanced Load \& Computation}
\label{sec:static-limitation}

\smallskip\noindent\textbf{Data Reuse.} 
The scheduling order of $M$, $N$, and $K$ determines the reuse of $A$, $B$, and $C$. 
Input reuse and output reuse are both governed by where $K$ sits in the loop nest. The further outward $K$ is placed (relative to $M$ or $N$), the more reuse of $B$ or $A$, respectively. Conversely, the further inward $K$ is placed, the more reuse of $C$.

For \textit{inner product}, $K$ is in the innermost loop; therefore, only outputs are reused, as shown in a single iteration in Fig.~\ref{fig:dataflow-bg}(a). However, there is possible data reuse of $A$ across iterations if we consider $A$ to be stationary. 
For \textit{outer product}, $K$ is in the outermost loop, so both $A$ and $B$ exhibit good data reuse. However, each iteration generates an entire $T_{M,N,k}$, 
meaning that revisiting the same output entry is distant in the loop nest. The upper bound (dense) distance between two outputs that must be accumulated is $M\times N$, and this distance varies based on the occupancy of $A$ and $B$.

For \textit{Gustavson}, $K$ is in the middle loop and $N$ is in the innermost loop. This allows $A$ to be fully reused. The output reuse is less distant than in \textit{outer product}, since each iteration generates only a row of the partial sum. The upper bound (dense) distance between two accumulated outputs is $N$. However, this dataflow sacrifices the potential for good reuse of $B$ and still faces the variable intermediate output sizes. Therefore, no single static schedule can simultaneously maximize reuse of all inputs and outputs.

\smallskip\noindent\textbf{Imbalanced Load \& Computation.} Static scheduling exacerbates imbalance in load and computation, especially when there is non-uniformity in the data.
For \textit{inner product}, static scheduling always computes between a row vector and a column vector. The actual number of intersections is determined by the nonzero positions in the two vectors. As shown in Fig.~\ref{fig:dataflow-bg}(a), the number of multiplications and additions varies across iterations. For \textit{outer product}, static scheduling causes each iteration to generate an entire matrix as a partial sum. The number of multiplications depends on the nonzeros in the column and row. The number of additions depends entirely on the positions of sparse values in $A$ and $B$. As shown in Fig.~\ref{fig:dataflow-bg}(b), there is only one addition across the matrix, while other positions have zero additions.
For \textit{Gustavson}, the computation imbalance is similar to that of \textit{outer product}. Additionally, static scheduling causes load imbalance across $A$ rows, which further harms the reuse of $B$ rows, as shown in Fig.~\ref{fig:dataflow-bg}(c).  

\subsection{Motivation for Dynamic Dataflow}
Static scheduling fixes the execution order and resource assignment ahead of time, leaving it unable to react to variations in nonzero distribution or rebalance work at runtime. As a result, no static dataflow can simultaneously maximize reuse on all three operands. We propose a dataflow that integrates \emph{dynamic scheduling} to adapt work selection within a tile and \emph{dynamic mapping} to redistribute partial sums across PEs at runtime, broadening the achievable reuse--utilization tradeoff.

\section{Segment Dataflow}
Compared to prior dataflows that follow a fixed loop order and iterate strictly over static loop bounds, we propose a new dynamic dataflow, which we call \emph{Segment}. Segment dataflow uses an active window to reorganize iteration over $A_{m,k}$ to maximize reuse and load balance. To enable efficient management of irregular intersections, we maintain partial results in a compressed, ordered coordinate space mapped to PEs. We use the term \emph{segment} because it represents the bounded path that an element takes through this coordinate space from injection to accumulation.

\begin{figure*}[t]
    \centering
    \includegraphics[width=0.96\linewidth]{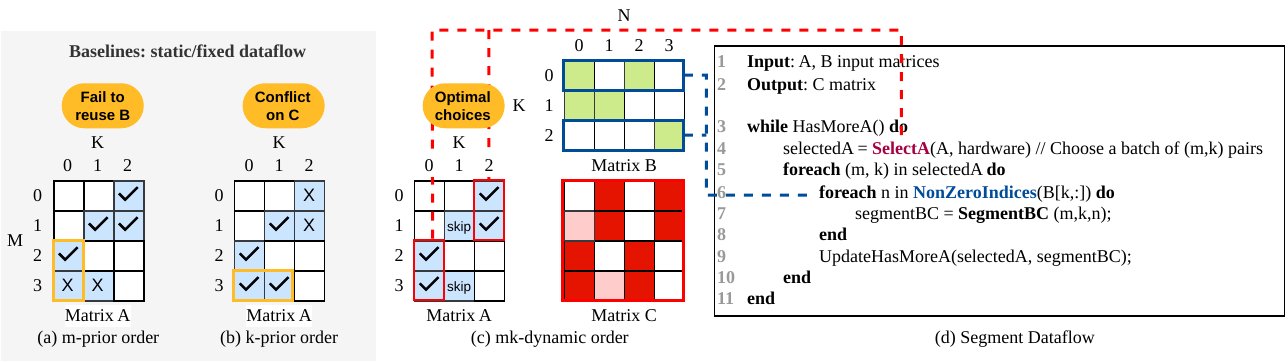}
    \vspace{-0.1in}
    \caption{SegFold Dataflow Example.  A batch of \((m,k)\) pairs is chosen by \textsc{SelectA}, and \textsc{SegmentBC}\((m,k,n)\) is invoked
    over \(B[k,:]\) nonzeros for ordered processing in the virtual coordinate space. Reuse benefits over two iterations are shown in pink.
    Tick marks indicate the \(A\) elements selected for each spatial iteration under different dataflows.
    }
    \label{fig:segfold-dataflow-example}
    \vspace{-0.15in}
\end{figure*}

Figure~\ref{fig:segfold-dataflow-example}(d) summarizes the algorithm. It introduces two key components that add dynamism:

\begin{enumerate}
    \item \textsc{SelectA}: dynamically selects and reorders $(m,k)$ pairs, breaking static $m$–$k$ nesting to increase $B$ reuse and spatial parallelism, while allowing partial $B$-row processing to absorb row-wise irregularity. Each invocation produces up to $R_{\max}$ (the PE-row capacity) pairs, where multiple pairs may share the same $k$ to enable $B$-row reuse but no two share the same $m$ to avoid $C$-row contention.
    \item \textsc{SegmentBC}: performs on-the-fly $B$–$C$ intersections across segments, exploiting spatial/temporal locality in $C$ to improve reuse of $C$ and reduce irregular reduction overhead. It consumes the selected pairs and their partial $B$ rows from \textsc{SelectA} and routes each $B_{k,n}$ element to find or create its position in a coordinate space of $C$.
\end{enumerate}

The main advantage of Segment dataflow is its ability to simultaneously exploit reuse across all three operands while increasing parallelism by combining multiple granularities within a unified execution model. Specifically, it captures element-wise reuse of $A$, row-wise reuse of $B$ across different $A$ elements, and tensor-wise reuse of $C$ across different partial-sum iterations. In addition, Segment dataflow provides element-wise flexibility in $C$ reductions by allowing elements within the tensor to move dynamically, thereby redistributing the reductions to allow parallelization. Canonical sparse dataflows do not provide this combination of properties. IP only exploits element-wise reuse of $C$; OP enables row-wise reuse of $A$ and $B$ but suffers from poor $C$ reuse and high irregular reductions; Gustavson achieves element-wise reuse of $A$ and row-wise reuse of $C$, but misses opportunities for reusing $B$ and still incurs significant reduction irregularities on $C$.

SegFold realizes this dataflow through \textsc{SelectA} and \textsc{SegmentBC}, as illustrated in Fig.~\ref{fig:segfold-dataflow-example}(d). Matrix $A$ is processed at element-wise granularity, while matrix $B$ is processed at row-wise granularity. In each iteration, \textsc{SelectA} examines the active window over $A$ and greedily selects $(m,k)$ pairs that maximize row-wise reuse of $B$. The selected $A$ elements are then broadcast together with their corresponding $B$ rows to exploit element-wise reuse. For each $B_{k,n}$ element, \textsc{SegmentBC} dynamically routes it within a row of the tensor-wise $C$, either accumulating it into an existing partial sum or inserting a new entry. Across consecutive iterations, reductions for $C$ elements are redistributed dynamically to mitigate register pressure and improve load balance.

\subsection{\textsc{SelectA}: row-wise intersection with reordering $K$}
\label{sec:selectA}

Because matrix $B$ is processed at row-wise granularity, dimension $N$ is placed in the innermost loop, as shown in Fig.~\ref{fig:segfold-dataflow-example}(d). This inherently maximizes the reuse of $A$. The nonzero positions of $A$ determine which rows of $B$ are accessed in each iteration. \textsc{SelectA} therefore picks a set of $A$ elements that maximizes \textbf{row-wise intersection}---the number of $m$ indices sharing the same $k$---which directly increases reuse of the corresponding $B$ row.

\begin{algorithm}[t]
\small
\SetAlgoLined
\LinesNumbered
\caption{\textsc{SelectA}: Dynamic $(m,k)$ Selection}
\label{alg:selectA}
\KwIn{$A$ bitmask, $B$-row metadata, hardware state; $W_{\max}$ window size; $R_{\max}$ PE-row capacity}
\KwOut{Selected $(m,k)$ pairs, corresponding partial $B$ rows}

\tcp{Inter-tile: sliding window over $K$}
\While{$|W_k| < W_{\max}$ \textbf{and} \textsc{HasMoreK}()}{
    $W_k \gets W_k \cup \{\text{next } k\}$\;
}

\tcp{Intra-tile: greedy $mk$-dynamic selection}
$selected \gets \emptyset$;\quad $usedM \gets \emptyset$\;
\ForEach{$k \in W_k$}{
    \lIf{$|selected| \geq R_{\max}$}{\textbf{break}}
    \ForEach{\textbf{parallel} $m$ \textbf{s.t.} $A[m,k] = 1$}{
        \If{$m \notin usedM$ \textbf{and} $|selected| < R_{\max}$}{
            $selected \gets selected \cup \{(m,k)\}$\;
            $usedM \gets usedM \cup \{m\}$\;
        }
    }
}

\tcp{Inter-tile: Retire completed $k$s and refill window}
\ForEach{$k \in W_k$ \textbf{s.t.} \textsc{AllDone}$(k)$}{
    $W_k \gets W_k \setminus \{k\}$\;
}

\Return $selected$\;
\end{algorithm}

Algorithm~\ref{alg:selectA} formalizes \textsc{SelectA}: it selects $(m,k)$ pairs from the input metadata based on the runtime hardware state. The flexibility in \textsc{SelectA} originates from the associativity of the reduction over the $K$ dimension, which allows intersections in Eq.~\ref{eq:mul} to be computed in any order without violating the accumulation rule in Eq.~\ref{eq:add}. Based on this observation, \textsc{SelectA} performs two levels of reordering: intra-tile reordering, which maximizes the number of $A$ entries selected within each $k$ column, and inter-tile reordering, which maximizes the number of $k$ columns considered concurrently.

\smallskip\noindent\textbf{Intra-tile Reordering.} Unlike prior dataflows that follow either an $m$-prior or a $k$-prior order, \textsc{SelectA} decides the set of $(m,k)$ pairs using an $mk$-dynamic order. More specifically, we simultaneously consider the following two criteria. First, we greedily maximize the number of pairs that share the same $k$, as shown in Algorithm~\ref{alg:selectA} line 5. Second, we avoid selecting pairs with different $k$ values but the same $m$ index, as shown in Algorithm~\ref{alg:selectA} line 8, because they update the same output row and can create reduction conflicts when their generated partials share column positions.

We use the $A$ pattern in Fig.~\ref{fig:segfold-dataflow-example} to demonstrate the advantages of our algorithm over conventional ones. Gustavson dataflow (Fig.~\ref{fig:segfold-dataflow-example}(a)) follows an $m$-prior order and chooses $A_{0,2}$, $A_{1,1}$, $A_{2,0}$ and $A_{3,0}$. This restricts the potential reuse of the first $B$ row, corresponding to the two elements highlighted in the yellow box. Outer product (Fig.~\ref{fig:segfold-dataflow-example}(b)) follows a $k$-prior order and selects $A_{2,0}$, $A_{3,0}$, $A_{1,1}$ and $A_{3,1}$. Nevertheless, $A_{3,0}$ and $A_{3,1}$ have the same $m$ index, so their products target the same output row and may contend when the corresponding $B$ rows contribute to overlapping $n$, which we do not know beforehand. Meanwhile, our algorithm selects $A_{0,2}$, $A_{1,2}$, $A_{2,0}$, and $A_{3,0}$, exploiting $B$ reuse while avoiding reduction conflicts in $C$.

\smallskip\noindent\textbf{Inter-tile Reordering.} Instead of choosing (m,k) pairs from a fixed tile, \textsc{SelectA} maintains a \emph{sliding window} over the $K$ dimension. A $k$ value in the window is marked as complete and replaced by a new one once all $A$--$B$ intersections for that $k$ have been processed. 

\begin{wrapfigure}{r}{0.45\linewidth}
    \centering
    \vspace{-0.15in}
    \includegraphics[width=\linewidth]{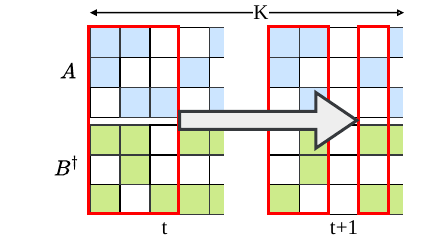}
    \vspace{-0.1in}
    \caption{Example active window of \textsc{SelectA}, shown in red for two time steps.}
    \vspace{-0.1in}
    \label{fig:segfold-window}
\end{wrapfigure}

For example, in Fig.~\ref{fig:segfold-window}, the third column of $A$ and the corresponding third $k$ slice of $B$ (shown transposed in the figure) each participate in only one effective intersection. Once that intersection completes, the window advances from $k=\{0,1,2\}$ to $k=\{0,1,3\}$. This enables $k$-level pipelining across different $(m,k)$ tiles. 

Moreover, using an entire $B$ row as the intersection granularity increases reuse of the triggering $A$ element across the loaded $B$ row, but it also reduces the available parallelism across different $B$ rows, because it requires the dataflow to finish an entire row of length $N$ before moving on to the next intersection.  To alleviate this limitation, we relax the requirement of fully loading a $B$ row before processing and instead allow the hardware to operate on \emph{partial} $B$ rows. Concretely, we decompose the $N$ dimension into smaller segments and interleave segments from multiple $B$ rows. A single active window tracks the progress of each $B$ row within the window, enabling higher inter-row parallelism while preserving the reuse benefits of row-wise intersection.

\subsection{\textsc{SegmentBC}: on-the-fly element-wise redistribution}
\label{sec:segmentBC}
The sparse patterns of both $T$ and $C$ are highly irregular and data-dependent, making it impractical to materialize a large intermediate tensor $T$ and then reduce it afterward. Therefore, Segment dataflow performs on-the-fly element-wise redistribution to (1) generate and maintain compressed indices for $C$ and (2) reduce each $T_{m,n,k}$ in place into the partial sum $C^*_{m,n}$\footnote{Here we use * to denote the intermediate result accumulated so far.}.

Let $(x,y)\in X\times Y$ denote the virtual coordinate at which a nonzero element of $C$ is allocated for reduction, where $|X|$ is the number of non-empty rows in $C$ and $|Y|$ is the maximum number of nonzeros in any row of $C$. We refer to $\mathcal{V}=X\times Y$ as the \emph{virtual coordinate space}; each occupied virtual coordinate holds a distinct $C$ element. 
As new $T_{m,n,k}$ are deposited into $C^*_{m,n}$, the virtual coordinate space evolves over time: newly created $C$ entries are inserted, shifting existing entries to maintain ordering.
In the $\mathcal{V}$ space and the algorithm description, we do not distinguish $T$ from $B$, since the dynamics depend only on the $n$ indices.
We use $f_t$ to denote the mapping from the dense Cartesian space to the $\mathcal{V}$ space at time step $t$:
\begin{equation}
    f_t(m,n)=(x,y).
\end{equation}
$f_t(m,n)$ indicates the compressed virtual location that stores the partial sum $C^*_{m,n}$ at time $t$.

To preserve both correctness and maximize locality, we constrain the mapping $f_t$ with the following properties:
\begin{enumerate}
    \item \textbf{Injectivity.} $f_t$ is injective: partial sums for distinct $(m,n)$ pairs are assigned to distinct virtual coordinates $(x,y)$. 
    \item \textbf{Row saturation.} Within each virtual row, nonzeros occupy consecutive $y$ positions, left to right (no gaps).  
    \item \textbf{Column ordering.} Within each virtual row, column indices strictly increase from left to right. That is, $\forall m$, if $f_t(m,n_1) = (x,y_1)$ and $f_t(m,n_2) = (x,y_2)$ with $n_1 < n_2$, then $y_1 < y_2$. 
    \item \textbf{Time ascending.} 
    Entries in a virtual row only move ``forward'' over time. For each $x$, if $f_t(m,n) = (x,y)$ and $f_{t'}(m,n) = (x,y')$ with $t < t'$, then $y \leq y'$. 
\end{enumerate}

\begin{figure}
    \centering
    \includegraphics[width=0.95\linewidth]{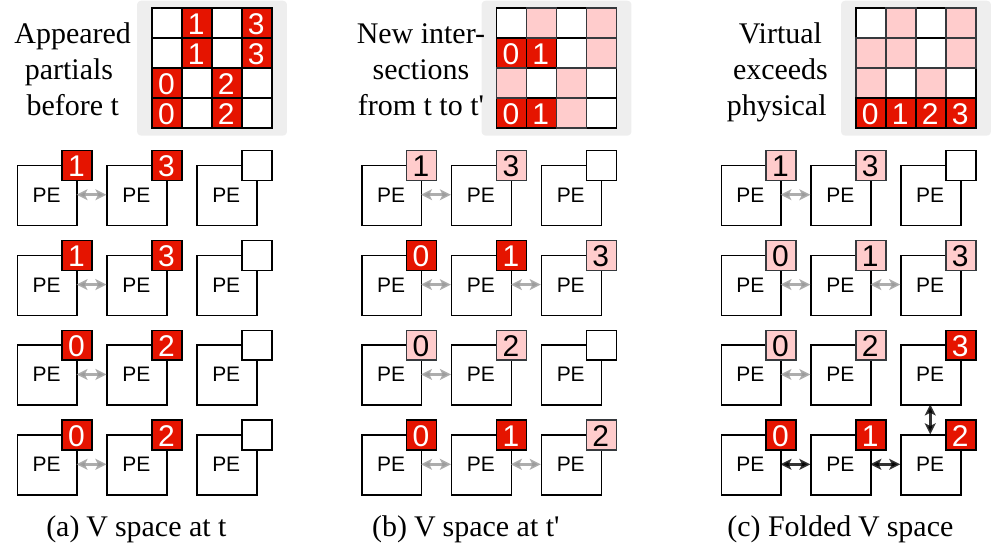}
    \caption{Example of on-the-fly update of the mapping from the $C$ column indices to the $\mathcal{V}$ space. Example from Figure~\ref{fig:segfold-dataflow-example}(c).}
    \vspace{-0.1in}
    \label{fig:segfold-segmentBC}
\end{figure}

Fig.~\ref{fig:segfold-segmentBC} illustrates how the mapping $f$ is updated in the walk-through example from Fig.~\ref{fig:segfold-dataflow-example}(c). The numbers in the boxes show the column index $n$ in the dense Cartesian space. At time step $t$, the bottom row of $f_t$ contains $\{0, 2\}$ (Fig.~\ref{fig:segfold-segmentBC}(a)). When new $B$ elements arrive and intersect with matching $A$ elements, they generate partial sums with column indices $\{0,1\}$. Because the partial sum with column index 1 has not previously appeared in the $\mathcal{V}$ space, the mapping is updated to $\{0, 1, 2\}$ to satisfy the column-ordering property (Fig.~\ref{fig:segfold-segmentBC}(b)). 

\reve[E-Q1]{With the definition of the $\mathcal{V}$ space, we formally define \emph{segment} as the displacement, measured in virtual coordinate positions, that a $B$ element traverses before reaching its final position in the PE array.} The name \textsc{SegmentBC} reflects that an element enters the $\mathcal{V}$ space as $B$ carrying $B$'s metadata and leaves as $C$ with its value and metadata updated. Specifically, for a $B$ element with column index $n$ that enters the array and is eventually accumulated into row $m$ of $C$:
\begin{equation}
    \text{displacement} = \|f_{t_{\text{in}}}(m,n) - f_{t_{\text{out}}}(m,n)\|,
\end{equation}
where $t_{\text{in}}$ and $t_{\text{out}}$ denote the times when the element enters and is consumed, respectively. \reve[E-Q1]{The displacement arises because the $\mathcal{V}$ space mapping $f$ evolves over time: as new $A$--$B$ intersections are discovered on-the-fly, newly formed $C^*$ entries are inserted into the virtual coordinate space, shifting existing entries and potentially increasing the distance from $f_{t_{\text{in}}}$ to the final position $f_{t_{\text{out}}}$. Additionally, the initial mapping $f_{t_{\text{in}}}$ may be approximate due to hardware constraints ($\S$\ref{sec:lut}).}

The optimization goal of \textsc{SegmentBC} is to minimize segment displacement, since longer displacements cause more network contention.
We employ a dynamic mapping strategy that determines where $B$ elements are placed and how they traverse the array; we discuss data traversal in $\S$\ref{sec:merge-network} and mapping in $\S$\ref{sec:lut}.

\section{SegFold Microarchitecture}

\begin{figure*}
    \centering
    \includegraphics[width=0.8\linewidth]{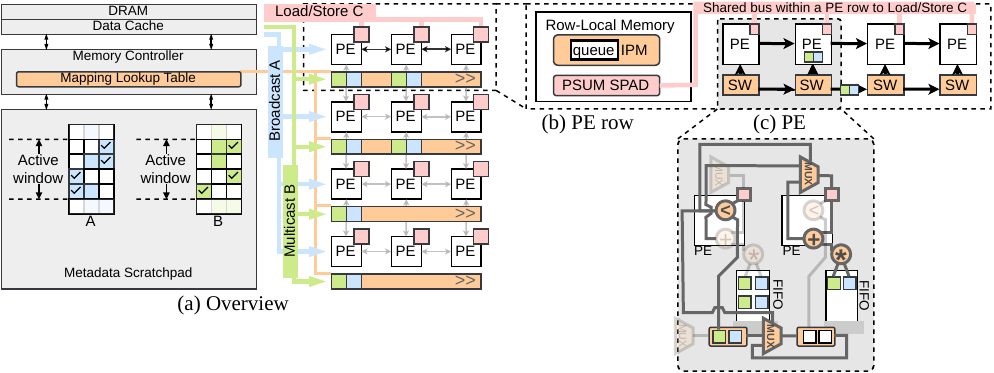}
    \centering
    \caption{SegFold $\mu$arch overview. PEs communicate over a local network, and share a row-level memory.}
    \vspace{-0.1in}
    \label{fig:segfold-arch}
\end{figure*}

Figure~\ref{fig:segfold-arch} presents a high-level overview of the SegFold microarchitecture. SegFold adopts a two-dimensional array of processing elements (PEs), following the architectural principles of prior spatial accelerators \cite{gpu, tpu, plasticine, trapezoid}. In terms of the stationarity taxonomy~\cite{teaal}, SegFold is effectively \emph{row-stationary} at the PE-row granularity, where each row owns an entire virtual row of $C$ for its lifetime, but only \emph{approximately} output-stationary at the PE level, since outputs may be remapped between PEs by spatial folding or spad spills. This two-level view captures SegFold's tradeoff: strict stationarity at the finest grain is sacrificed for the flexibility to rebalance load at runtime.

To efficiently orchestrate data movement, SegFold integrates a hybrid on-chip interconnect.
At the single-PE level, each PE communicates with its nearest neighbors, forming a mesh network across the 2D PE array. At the PE-row level, each PE row has a dedicated \emph{row shifter} that shifts and streams the partial $B$ row horizontally into the PEs. Across PE rows, a \emph{vector multicast network} routes $B$ segments to the appropriate PE rows, enabling flexible and conflict-free global redistribution. Together, these components provide both fine-grained local communication and high-bandwidth global redistribution essential for SegFold's dynamic dataflow.

To enable high intra-tile reuse, SegFold employs an on-chip cache to store tiles of $A$ and $B$. $\S$\ref{sec:memory-controller} describes the memory controller, which is the key component realizing the Segment dataflow. As shown in Figure~\ref{fig:segfold-arch}, this unit is connected to the on-chip metadata buffer and to the Index-to-PE Mapper (IPM)---a lookup table that maps each incoming $B$ element's column index to its starting PE position in the merge network ($\S$\ref{sec:lut}), enabling it to perform the \textsc{SelectA} scheduling algorithm and \textsc{SegmentBC} mapping algorithm respectively. It ensures that the PE array will receive a steady, reuse-optimized supply of matrix tiles and minimizes off-chip memory accesses.


\subsection{PE Row}
\label{sec:pe-row}

Figure~\ref{fig:segfold-arch}(b) illustrates a representative PE row containing four PEs. Each PE integrates an arithmetic logic unit (ALU) for local computation, a FIFO buffer that stores matched-but-not-yet-consumed $B$ values, and a router with four physical ports to its neighbors. 
Although the router supports four-way connectivity, only one direction is active at any time, determined by the folding mechanism ($\S$\ref{sec:folding}).

Initially, the router's active direction is \emph{rightwards}, consistent with the default column-ordering rule of the $\mathcal{V}$ space mapping. Here, the PE row behaves as a simple right-propagating merge network, forming the baseline for our design. The folding technique ($\S$\ref{sec:folding}) modifies these active directions.

\revf[F-Q2]{
We first introduce our PE-to-Virtual-Coordinate mapping. Each PE in a row holds exactly one virtual coordinate position of the output matrix $C$. Concretely, PE $p$ in row $r$ stores the partial sum $C^*_{m,n}$ whose $\mathcal{V}$ space coordinate is $(r, p)$, along with the Cartesian column index $n$ that maps to this position. A PE row therefore represents one virtual row of $C$, with PEs ordered left-to-right by increasing column index. When a $B$ element with column index $b$ enters a PE row, it is injected at the starting position determined by the IPM ($\S$\ref{sec:lut}) and traverses rightward through the merge network, comparing $b$ against each PE's stored column index $c$ until it finds a match ($b = c$, triggering accumulation) or a gap ($b < c$, triggering insertion).}

Each PE row also has a dedicated connection to a small local memory containing: (i) the IPM, which the merge network keeps up-to-date as $C$ entries shift, and (ii) a scratchpad (spad) holding overflow $C$ values. All PEs within a row share this memory interface; therefore, we limit accesses to avoid contention.

\subsubsection{Adaptive Merge Network}
\label{sec:merge-network}

We refer to the PEs' local interconnect as a \emph{merge network}. Each PE contains a \emph{merger} (the local component of the merge network) that compares the incoming $B$ element's column index against the $C$-column index stored at this $\mathcal{V}$ space position. Based on this comparison, the merger decides whether to forward $b$ to a neighbor or retain it locally for accumulation.

\begin{figure}[t]
    \centering
    \includegraphics[width=\linewidth]{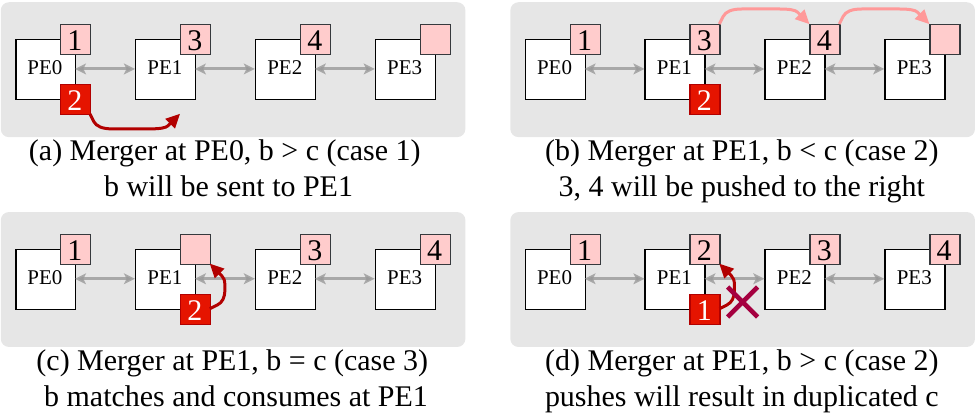}
    \vspace{-0.2in}
\caption{Example of on-the-fly intersection 
where the indices of $B$ are $>$, $<$, $=$, and conflicting with the $C$ index.}
    \label{fig:segment-example}
\end{figure}

Figure~\ref{fig:segment-example} illustrates how the merge network determines the final virtual position $f_{t_{\text{out}}}$ in \textsc{SegmentBC} for each incoming $B$ element. We use the red boxes to denote the Cartesian column index of the incoming $B$ element and the pink boxes to denote the Cartesian column index of the $C^*$ element stored at the PE. 

Consider a PE row where the $C^*$-column indices $\{1, 3, 4\}$ are stored at positions $y = \{0, 1, 2\}$. A $B$ element with column index $b = 2$ arrives at $f_{t_{\text{in}}} = 0$. At each position, the merger compares $b$ and $c$ (we use lower case to represent column indices), leading to one of these cases:

\begin{itemize}
    \item \textbf{Fig.~\ref{fig:segment-example}(a): $b\!>\!c$.}  
    Because the $C^*$-column indices are strictly monotonic, the final position $f_{t_{\text{out}}}$ must lie to the right of $y$.  
    The element is therefore forwarded to $y\!+\!1$.

    \item \textbf{Fig.~\ref{fig:segment-example}(b): $b\!<\!c$.}  
    If we eventually reach a position $y$ with a larger $c$, because monotonicity ensures that no match exists further to the right, the network shifts all $C^*$-column indices to the right by one slot.
    This creates an empty slot at $y$: the final virtual position $f_{t_{\text{out}}}$.

    \item \textbf{Fig.~\ref{fig:segment-example}(c): $b\!=\!c$.}  
    A match is found at position $y$, which becomes $f_{t_{\text{out}}}$.
\end{itemize}

Together, these cases ensure that each $B$ element successfully finds its final virtual position $f_{t_{\text{out}}}$, provided that the $C^*$-column indices to the left of $f_{t_{\text{in}}}$ satisfy $c < b$. 
Figure~\ref{fig:segment-example}(d) shows a scenario that violates this, and is prohibited by our dataflow.

\subsubsection{Index to PE Mapper (IPM)}
\label{sec:lut}

To avoid the prohibited scenario in Fig.~\ref{fig:segment-example}(d), the key challenge is to construct a mapping that (i) guarantees a legal $f_{t_{\text{in}}}$ for every segment and (ii) simultaneously minimizes the traversal distance required. Achieving both goals increases PE utilization by enabling more pipelined segment injections from different $B$ rows, while also reducing merge-network latency and contention.

We will now use $s$ as shorthand for $f_{t_{\text{in}}}$ (where the
$B$ element enters the network). As discussed, a starting point $s$ is legal if all $C^*$-column indices that appear to the left of $s$ are strictly smaller than $b$, i.e., $\forall i < s,\; c_i < b$. Due to row saturation and the column-ordering property, the $c_i$ values stored in the merges are guaranteed to be strictly ordered from left to right without gaps. This monotonicity allows us to perform a binary search over the $c_i$ values to determine the rightmost legal starting point for $b$. 

\begin{figure}
    \centering
    \includegraphics[width=\linewidth]{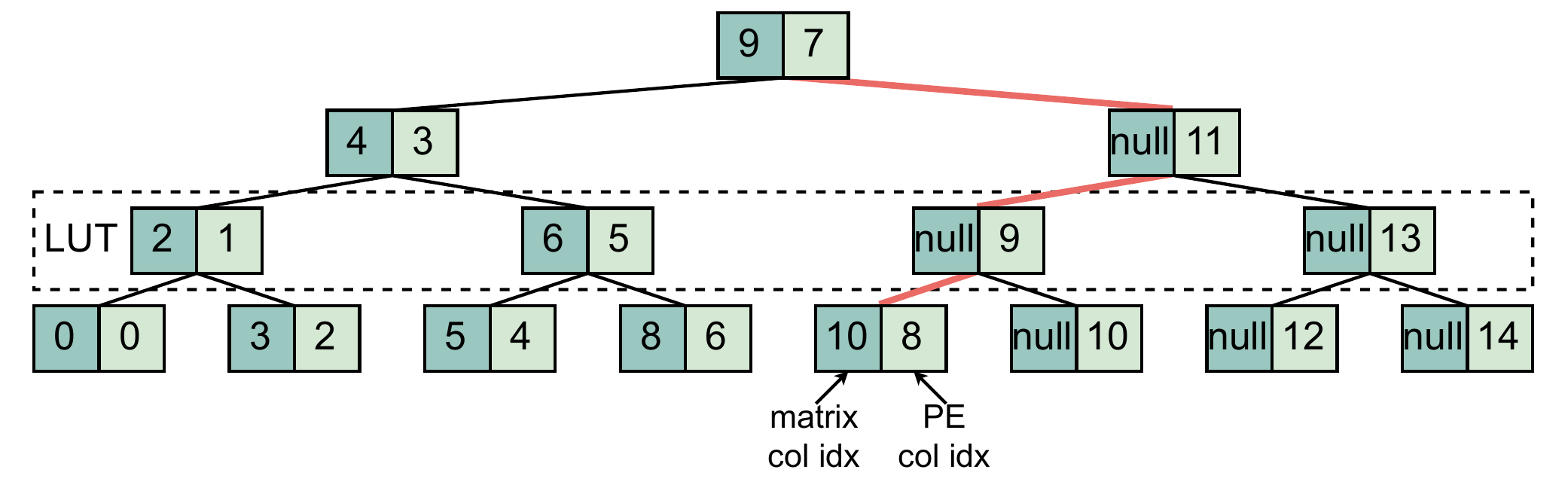}
    \vspace{-0.25in}
    \caption{Index to PE mapper (IPM) using binary search.
    Each level is in its own look up table (LUT) for pipelining.}
    \label{fig:micro-lut}
\end{figure}
Figure~\ref{fig:micro-lut} shows an example of an IPM with multiple lookup tables (LUTs), looking up a $B$ element with column index 11. In the first level, it is greater than 9 and will go to the right branch. In the second and third level, as both keys are showing null, it will go to the left branch. Finally, it reaches a leaf with value 8. Therefore, this $b$ element will be mapped to the 8th merger and traverse the adaptive merge network.

\smallskip\noindent\textbf{Row-wise Mapping.}
Computing a separate mapping for every $b$ element in a $B$ row would be prohibitively expensive in hardware, since it would require the IPM to support a number of simultaneous reads proportional to the number of PEs. To reduce this overhead, we instead compute the mapping only for the \emph{first} nonzero element of each $B$ row. This optimization is valid because, under the row mapping, each element's start point $s$ increments by 1, matching the increment of its column index $b$. Combined with the column-ordering property, this guarantees that if the first mapped element of a row satisfies the legality condition $\forall i < s,\; c_i < b$, the remaining elements automatically satisfy it as well. 

\smallskip\noindent\textbf{IPM Updates.}
When a PE updates its $c$ value, it notifies the IPM. Because the LUT has a limited number of write ports, multiple updates to the same entry are queued and applied serially, so the IPM may not always reflect the most recent $c$ value. Due to the time-ascending property of these updates, an out-of-date LUT can only map a $b$ element to a position to the \emph{left} of its true most-recent legal starting point. This does not violate correctness, since the merge network still identifies the correct match, but it may increase the segment length. We quantify the gap between fully up-to-date $c$ indices and the LUT-based mapping with bounded write bandwidth in $\S$\ref{sec:ablation}.

\subsection{Memory Controller}
\label{sec:memory-controller}
The memory controller in SegFold realizes the scheduling logic of the Segment dataflow: it maintains an active window over $B$ rows, filters rows by intersection, and tracks partial-row progress, while also serving the underlying load/store requests for $A$, $B$, and $C$. Since the goal of \textsc{SelectA} is to choose multiple $A$ values from the same column, $A$ is stored in a column-major format. In contrast, $B$ is processed at row granularity, so it is stored in a row-major format. Only nonzero elements of both matrices are stored. Because accesses typically touch consecutive elements within a column of $A$ or a row of $B$, the memory controller includes a coalescing unit that merges fine-grain requests before issuing them to the cache and DRAM.

As discussed in $\S$\ref{sec:selectA}, the active window tracks a bounded set of $K$ values for both $A$ and $B$.
The metadata for $A$ is a compact bitmask recording consumption status. 
The \textsc{SelectA} logic scans this $A$-side bitmask over the active window each cycle to avoid $m$-index conflicts among already-selected pairs and to maximize the nonzero $k$-index within a single column. The bitmask read and window scan are $O(W)$ in the window size; for our default $W\!=\!32$, the entire scan completes within a single cycle as a modest combinational circuit.

For $B$, we use a doubly compressed sparse-row format (DCSR)~\cite{dcsr}, augmented with an extra start pointer per active row to track unprocessed elements. The second level of compression skips empty rows in $O(1)$ during scheduling, which is critical for highly sparse matrices where many rows in the active window contribute no nonzero intersections with the selected $A$ columns and would otherwise be enumerated explicitly.

Supporting intra-window reordering of $k$ means that a larger window enables more $B$ rows to be processed in parallel. However, a larger window increases metadata size and update traffic. In $\S$\ref{sec:sensitivity}, we evaluate different window sizes and select a configuration that preserves performance while capping metadata overhead.

\subsection{Vector Multicast Network}
\label{sec:vec_multicast}
SegFold employs \reva[A-Q3]{a vector multicast network~\cite{scalagraph}} for loading $B$ rows.
Each PE row is responsible for processing at most one $B$ row at a time. Because elements within a $B$ row must be mapped to a consecutive set of mergers—with the head position determined dynamically—we incorporate a row shifter that realigns the incoming $B$ elements to the appropriate consecutive positions.
Furthermore, \textsc{SelectA} may choose multiple distinct $k$ values simultaneously, enabling several $B$ rows to be active in the same cycle. As shown in Fig.~\ref{fig:segfold-arch}(a), the network for $B$ must multicast multiple $B$ rows into the PE rows; we therefore use a vectorized crossbar to provide the multicast bandwidth from the memory backend to the PE rows. Because streaming multiple vectors increases complexity at the network ports, a key design parameter is the number of $B$ rows that can be multicast in parallel. Our design allows up to 4 vector multicasts, which balances area/energy efficiency and performance.

\subsection{Folding: Mapping Segments to PEs in Space and Time}
\label{sec:folding}

The final $\mathcal{V}$ space representation of $C$ will exhibit highly irregular row lengths, and each virtual position may require a different number of MAC operations. To maximize utilization and bandwidth, we do not constrain the $\mathcal{V}$ space row length to be less than the physical length of a PE row. Instead, we introduce two complementary techniques—\emph{spatial folding} and \emph{temporal folding}—that effectively map an irregular $\mathcal{V}$ space onto a regular 2D PE array. 

\smallskip\noindent\textbf{Spatial folding} addresses the mismatch between the irregular, variable-length rows in $\mathcal{V}$ space and the fixed physical width of the PE array. A long virtual row of $C$ may extend beyond the capacity of a single PE row, while a short row may leave many PEs idle. To avoid underutilization, SegFold allows each logical row of outputs to \emph{fold} across multiple PE rows.

Consider a PE array with $R$ rows and $P$ columns, indexed by coordinates $(r,p)$ for
$0\!\le r<R,\;0\!\le p<P$. 
Each router maintains a direction configuration, which specifies which
of the four output ports will be used to place the next logical column of the $\mathcal{V}$ space row.
When an on-the-fly intersection generates a new virtual column $j$ in an occupied physical position $(r,p)$,
the router at $(r,p)$ examines its four neighboring PEs in the following priority: \{right, up, down, and left\}. Right is always the first priority to first realize the default merge network configurations. 

Formally, it selects the first coordinate
\begin{equation}
(r',p') \in \bigl\{
(r,p+1),\ (r-1,p),\ (r+1,p),\ (r,p-1)
\bigr\}
\end{equation}
whose occupancy bit satisfies $\mathcal O_{r',p'} = 0$.
If such a neighbor exists, the router sets $\mathcal O_{r',p'} \leftarrow 1$, forwards the
virtual column $j$ to that PE, and updates its direction configuration to point to $(r',p')$
for the next placement. 
In the example of Fig.~\ref{fig:segfold-segmentBC}(c), the fourth element with index 3 in the bottom row is folded to the upper PE with an updated router configuration. 
Because each router exposes only its highest-priority free neighbor
as the active direction, the spatial footprint of the $\mathcal{V}$ space row grows smoothly: horizontally when possible, and folding vertically when the
row exceeds the local PE-row capacity. 
Through this mechanism, spatial folding enables irregular, long $\mathcal{V}$ space rows to occupy additional PE rows when necessary, while allowing shorter rows to leave unused PEs available for other work, thereby improving overall array utilization.

\smallskip\noindent\textbf{Temporal folding} handles overflow and imbalance in the $C$ reductions. While spatial folding effectively manages irregular row lengths and maps them onto the regular PE mesh, it does not guarantee that the number of nonzeros in a virtual row is always bounded by the total PE-array capacity. To support rows that exceed this capacity, each PE row is equipped with a spad that stores overflow virtual rows.

When a PE needs to accommodate a new partial sum, it can overwrite its local $c_i$ register in place and spill the previous partial sum into the spad. This mechanism not only supports rows that are longer than the physical array, but also helps mitigate reduction imbalance: partial sums that are likely to complete early can be offloaded to the spad, allowing PEs to be reused for other work while preserving correctness.

\revb[B-Q1]{
\subsection{Scalability}
SegFold's control complexity scales linearly with array size in the dominant terms. The merge network traversal within each PE row is $O(P)$ in the worst case, where $P$ is the number of PEs per row; however, the expected traversal is much shorter because the IPM provides a near-optimal starting position via $O(\log P)$ binary search, whose latency could be amortized through pipelining. The IPM itself uses a tree-structured lookup table whose depth grows as $\log P$, and whose total storage scales as $O(P)$ entries per row. The spad for overflow $C$ values scales linearly with the number of PE rows ($R$), as each row maintains an independent spad. Overall, scaling the array from $P$ to $2P$ PEs per row doubles the merge network width and IPM size, but does not change the asymptotic complexity of the control logic.}
\section{Methodology}

\smallskip\noindent\textbf{Simulation Infrastructure.}
We evaluate SegFold using a cycle-level microarchitectural simulator. Hardware parameters are summarized in Table~\ref{tab:config}. To maintain a hardware cost comparable to state-of-the-art sparse accelerators~\cite{flexagon}, SegFold instantiates a $16\times16$ 2D PE array. \revc[C-Q7]{The design meets timing at 1\,GHz.} The memory controller uses a fixed active-window size of 32, chosen to balance performance against metadata and storage overhead.

\begin{table}[t]
\caption{SegFold Hardware Configuration}
\vspace{-0.05in}
\setlength{\tabcolsep}{4pt} 
\footnotesize
\centering
\begin{tabular}{cc}
\toprule
\textbf{Parameter} & \textbf{Value} \\
\midrule
PE array & $16 \times 16$ \\
Per-row hardware ($\times 16$ rows) & shifter, spad bank, LUT bank \\
Active $B$ window size & 32 \\
Cache size, associativity, line size & 1.5\,MiB, 16-way, 128\,B \\
DRAM & HBM2-8Gb, 2\,Gbps \\
\bottomrule
\end{tabular}
\vspace{-0.15in}
\label{tab:config}
\end{table}

The memory hierarchy uses an on-chip cache backed by off-chip HBM2 DRAM, with details in Table~\ref{tab:config}.
Off-chip memory is modeled using Ramulator2~\cite{ramulator2} configured with an HBM2 at 2\,Gbps.
All hardware components are simulated on a cycle-by-cycle basis to ensure timing accuracy.  

\smallskip\noindent\textbf{Baselines.}
We compare to two state-of-the-art accelerators.

\emph{Flexagon}~\cite{flexagon} is a reconfigurable accelerator capable of supporting multiple canonical dataflows.
The on-chip components of Flexagon are modeled using the open-source \texttt{STONNE} simulator~\cite{stonne}, \reva[A-Q4]{integrated with the same Ramulator-based memory backend for consistency.}
Since Flexagon was originally designed as a 1D array accelerator, it is extended to a 2D configuration for fair comparison.
This extension duplicates the on-chip cache while maintaining a shared off-chip DRAM.
To match the compute resources of SegFold, Flexagon's 1D array of 128 PEs is scaled into a 2D array of $2 \times 128$ PEs.
On-chip bandwidth is preserved by scaling both the reduction network and distribution network to 128 elements per cycle for each 1D array. The original Flexagon network remains unchanged along the second dimension; the 2D extension is achieved by tiling along $M$ to distribute the workload evenly across the two PE arrays.

\emph{Spada}~\cite{spada} is a runtime-adaptive SpGEMM accelerator that dynamically adjusts its window to the sparsity of matrix $A$. We use its open-source simulator unchanged as our baseline. For the non-square evaluation (Fig.~\ref{fig:nonsquare-combined}(a)), we additionally integrate Spada with the same Ramulator2 HBM2 memory backend used by SegFold to ensure a fair memory-system comparison.

\smallskip\noindent\textbf{Area and Energy.} We employ the ASAP7 7nm standard-cell library as the target technology for RTL synthesis~\cite{clark2016asap7}. We use Design Compiler to elaborate all RTL sources and invoke \texttt{compile\_ultra} for timing-driven optimization. We report the post-synthesis timing, area, resources, and power.

\smallskip\noindent\textbf{Workloads.}
We select fifteen matrices from SuiteSparse~\cite{suitesparse} as our baseline benchmark suite for the overall and non-square comparisons, covering a range of application domains, matrix sizes, aspect ratios, and sparsity levels. The ablation studies in $\S$\ref{sec:ablation} draw from a slightly different subset to expose mechanism-specific behavior. Throughout, we report \emph{density} as $\mathrm{nnz}/(M\!\times\!N)$. These matrices are characterized in Table~\ref{tab:suitesparse}.
Across all experiments, the matrices span dimensions from hundreds to tens of thousands of rows and densities across more than two orders of magnitude. The test set includes both square and non-square matrices to capture the shape impact of matrix multiplication. The benchmark suite exercises a representative range of irregular access patterns. Unless otherwise specified, we use the transpose of the matrix as $B$ for matrix multiplication.

\begin{table}[t]
  \centering
  \footnotesize
  \caption{SuiteSparse matrices used in our evaluation.}
  \label{tab:suitesparse}
  \begin{tabular}{lrrrl}
    \toprule
    Matrix & $M$ & $N$ & Density & Application domain \\
    \midrule
    fv1            & 9604  & 9064  & $9.79\mathrm{e}{-4}$ & 2D/3D problem \\
    flowmeter0     & 9669  & 9669  & $7.21\mathrm{e}{-4}$ & Model reduction \\
    delaunay\_n13  & 8192  & 8192  & $7.32\mathrm{e}{-4}$ & Undirected graph \\
    ca-GrQc        & 5242  & 5242  & $1.05\mathrm{e}{-3}$ & Undirected graph \\
    ca-CondMat     & 23133 & 23133 & $3.49\mathrm{e}{-4}$ & Undirected graph \\
    poisson3Da     & 13514 & 13514 & $1.93\mathrm{e}{-3}$ & CFD \\
    bcspwr06       & 1454  & 1454  & $2.51\mathrm{e}{-3}$ & Power network \\
    tols4000       & 4000  & 4000  & $5.49\mathrm{e}{-4}$ & CFD \\
    rdb5000        & 5000  & 5000  & $1.18\mathrm{e}{-3}$ & CFD \\
    gemat1         & 4929  & 10595 & $8.92\mathrm{e}{-4}$ & Power network \\
    lp\_woodw      & 1098  & 8418  & $4.06\mathrm{e}{-3}$ & Linear programming \\
    pcb3000        & 3960  & 7732  & $1.88\mathrm{e}{-3}$ & Circuit simulation \\
    Franz6         & 7576  & 3016  & $1.99\mathrm{e}{-3}$ & Combinatorial problem \\
    Franz8         & 16728 & 7176  & $8.36\mathrm{e}{-4}$ & Combinatorial problem \\
    psse1          & 14318 & 11028 & $3.63\mathrm{e}{-4}$ & Power network \\
    \bottomrule
  \end{tabular}
\end{table}

\smallskip\noindent\textbf{Tiling.}  
Tiling is applied along the Cartesian dimensions $M$ and $N$. 
Tile sizes are statically determined based on the distribution of nonzero values in the corresponding $C$ tiles. 
To accommodate the tiling, when a virtual row of $C$ exceeds the physical PE-row capacity, overflow values are spilled to the per-row spad. Because the spad is sized to accommodate the expected maximum overflow, which is determined by the tile dimensions along $M$ and $N$ and the anticipated density of $C$, spills are infrequent under our default tiling configuration.

\section{Evaluation}
We evaluate SegFold's end-to-end performance against Spada and Flexagon, perform a per-component ablation, study sensitivity to key hardware parameters and input sparse patterns, and report post-synthesis area and power.

\subsection{Overall Performance}
\begin{figure*}[t]
    \centering
    \includegraphics[width=0.98\linewidth]{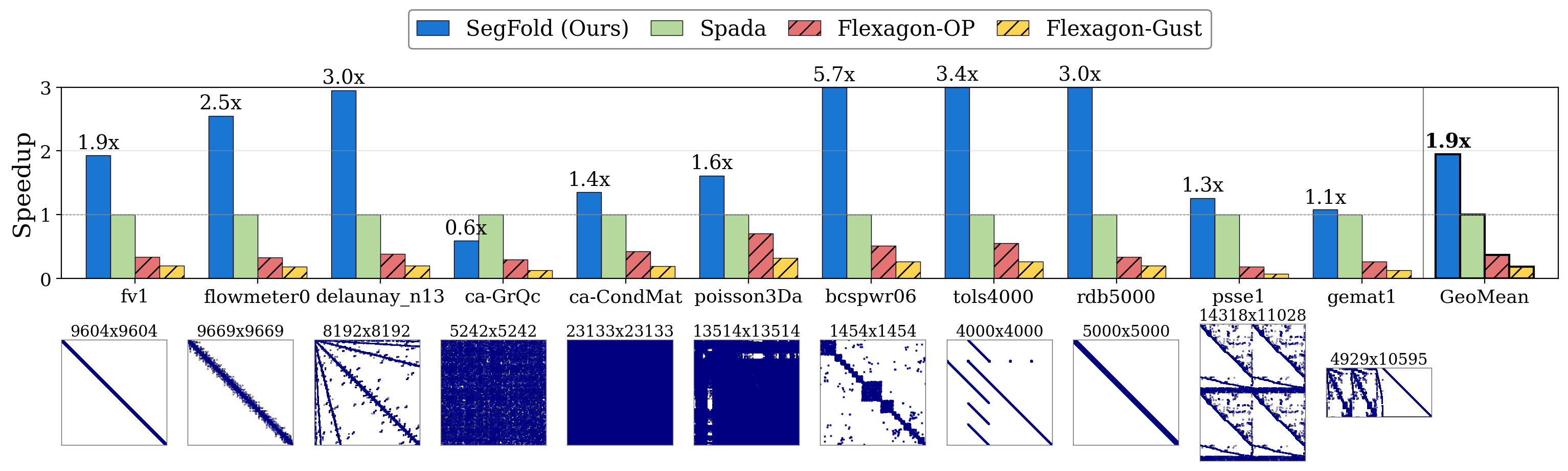}
    \vspace{-0.14in}
    \caption{Speedup over Spada and static dataflow implementations on SuiteSparse matrices. Sparsity patterns shown below.}
    \label{fig:suitesparse-speedup}
    \vspace{-0.1in}
\end{figure*}

\begin{figure}[t]
    \centering
    \includegraphics[width=\linewidth]{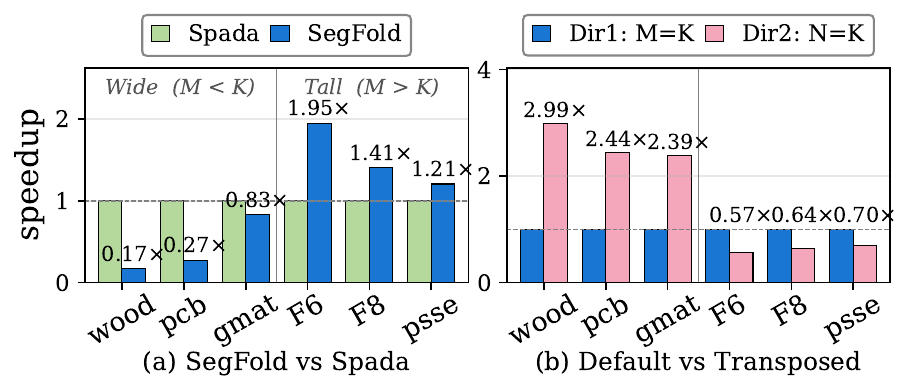}
    \vspace{-0.25in}
    \caption{Nonsquare SuiteSparse matrices.
        \textbf{(a)}~SegFold vs.~Spada speedup, normalized to Spada.
        \textbf{(b)}~Effect of multiplication direction on SegFold throughput,
        normalized to nonsquare matrix as $A$.}
    \label{fig:nonsquare-combined}
    \vspace{-0.09in}
\end{figure}

Fig.~\ref{fig:suitesparse-speedup} shows that SegFold achieves a $1.95\times$ geometric-mean speedup over Spada and a $5.3\times$ geometric-mean speedup over the best-performing Flexagon configuration across all workloads. The much larger $5.3\times$ speedup over Flexagon reflects the limitation of static dataflows: Flexagon must commit to a single dataflow---inner-product, outer-product, or Gustavson---per tile, and even its best per-tile choice cannot simultaneously exploit reuse on all three operands. SegFold's dynamic scheduling sidesteps this by reordering work within each tile based on the runtime nonzero pattern.

On highly sparse matrices with structured, non-uniform nonzero distributions, SegFold achieves $1.08\times$ to $5.75\times$ speedups over Spada. The advantage comes from SegFold's two core dynamic mechanisms operating \emph{within} a tile: (1) \textsc{SelectA} dynamically reorders $(m,k)$ pairs in the active window to maximize $B$-row reuse and avoid $C$-row contention, and (2) \textsc{SegmentBC} dynamically remaps partial sums across PEs based on the evolving $\mathcal{V}$ space state. Spada, in contrast, adapts only its window \emph{size} at tile granularity while keeping the scheduling within each window static, leaving sub-tile reuse and load-balance opportunities unexploited. This gap is amplified on highly sparse matrices, where SegFold's dynamic scheduling can skip whole empty regions of the iteration space outright (e.g., $k$ columns with no $A/B$-nonzeros), while Spada must still follow its static loop over the entire window even when most of it contributes no work. The one exception is ca-GrQc, on which SegFold underperforms Spada ($0.59\times$): its scale-free graph structure produces a few extremely dense rows that overwhelm SegFold's per-row PE allocation, while Spada's tile-level adaptation handles them better.

\subsection{Non-square Performance}
Figure~\ref{fig:nonsquare-combined}(a) compares SegFold against Spada on non-square SuiteSparse matrices. These non-square comparisons are done by multiplying each matrix by its own transpose. SegFold outperforms Spada on tall matrices, achieving $1.42\times$ geomean speedup over Spada, where its dynamic dataflow effectively exploits the elongated row structure. On wide matrices, however, SegFold falls behind: two out of three tested matrices underperform Spada. The cause is how SegFold handles the $K$ dimension: because we do not tile along $K$ within a tile, a large $K$ relative to $M$ creates load imbalance across $A$ rows. Transposing the matrices does not help here because we are already computing $A \times A^{\!\top}$ (a self-transpose multiply). However, there are cases where transposition could provide critical optimization. 

As shown in Fig.~\ref{fig:nonsquare-combined}(b), Direction~1 computes $A_\mathrm{real}^{M,K} \times S^{K,N}$, where $M\neq K$ and $K=N$ while Direction~2 computes $S^{M,K} \times \left(A_\mathrm{real}^{N, K}\right)^{\!\top}$, where $M=K$ and $K\neq N$. They are arithmetically equivalent up to an output transpose, effectively swapping which operand drives \textsc{SelectA}. Transposing the wide matrices recovers $2.4$--$3.0\times$ over Direction~1, confirming that the $M/K$ ratio has a significant impact on SegFold's performance: when an operand has its reduction-side dimension $K$ much larger than its output-side dimension, making it the second operand places its short axis along the multiplication's output dimension $N$. The dataflow then iterates along the short $N$ rather than the long $K$, improving the efficiency of \textsc{SelectA}'s scan. Conversely, on tall matrices Direction~1 is already favorable. Selecting the multiplication direction can often be decided in advance, yielding several-fold speedups.

\subsection{Ablation Studies}
\label{sec:ablation}
We isolate the effects of two components of SegFold: (i) the dynamic scheduling and (ii) the dynamic mapping.

\subsubsection{Effect of Dynamic Scheduling}
To understand the impact of SegFold's dynamic dataflow, we perform an experiment with fixed $k$ iteration order, making the dataflow resemble a constrained outer-product scheme: instead of dynamically reordering $k$ within the active window, we process $k$ in a predetermined sequence. The final result shows this reduces normalized performance to $0.670\pm0.065$ of the baseline, indicating that dynamic $k$ reordering is important for exposing segment-level parallelism and keeping PEs busy.

\subsubsection{Effect of Dynamic Mapping}
To isolate the effect of dynamic mapping, we compare SegFold's LUT-based mapping against two alternatives, both using the same dataflow and hardware resources but with different mapping logic:

\begin{enumerate}[leftmargin=*]
  \item \textbf{Zero-Offset mapping}: the head of the $B$ row is always mapped to the beginning of the PE row ($f_{t_{\text{in}}}(m,n)=0$).
  \item \textbf{Ideal-Network mapping}: an oracle mapper that always finds the optimal placement with no stale-index overhead.
\end{enumerate}

\begin{figure}
    \centering
    \includegraphics[width=1\linewidth]{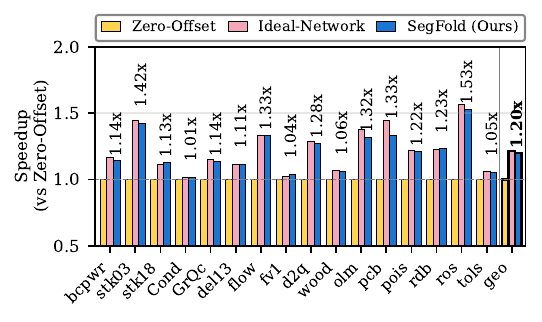}
    \vspace{-0.3in}
    \caption{Speedup of different mapping methods (Zero-Offset, Ideal-Network, SegFold) on SuiteSparse matrices, normalized to Zero-Offset.}
    \vspace{-0.14in}
    \label{fig:ablation-dynamic-scheduling}
\end{figure}

Figure~\ref{fig:ablation-dynamic-scheduling} reports the speedup of each mapping method across 16 SuiteSparse matrices, normalized to the zero-offset baseline. SegFold's LUT-based mapper achieves a geometric-mean speedup of $1.20\times$ over the zero-offset policy. Matrices with more irregular sparsity patterns (e.g., pcb3000, olm5000, flowmeter0) benefit most from dynamic scheduling, as the LUT-based mapper adapts to instantaneous PE occupancy and avoids long segment traversals. Compared to the ideal oracle mapping, SegFold's LUT-based scheduling incurs only a $1.2\%$ average overhead, demonstrating that our hardware approximation closely tracks the theoretical optimum.

\begin{figure}[t]
    \centering
    \includegraphics[width=\linewidth]{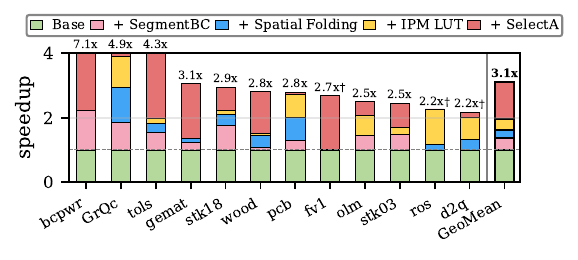}
    \vspace{-0.38in}
    \caption{Speedup break down of the different dynamic mechanisms.}
    \label{fig:ablation}
    \vspace{-0.09in}
\end{figure}

\revf[F]{
\smallskip\noindent\textbf{Attribution Summary.} To quantify the contribution of each dynamic mechanism in both the dataflow and microarchitecture, we conduct an incremental ablation study across 12 SuiteSparse matrices. Figure~\ref{fig:ablation} shows the performance breakdown of SegFold, achieving a geometric-mean speedup of $3.1\times$ over the base configuration. Across different sparsity patterns, each dynamic mechanism contributes to the overall speedup, with \textsc{SelectA} delivering the largest gain---indicating that dynamic scheduling plays a critical role in handling highly irregular sparse matrices. \textsc{SegmentBC}, spatial folding, and the IPM LUT provide additional gains, with their relative benefits depending on the matrix's structural properties and output sparsity patterns.}

\subsection{Sensitivity Studies}
\label{sec:sensitivity}
We study SegFold's sensitivity along two axes: \emph{hardware parameters} and \emph{input characteristics}. 

\subsubsection{Hardware Parameter Sensitivity}
\label{sec:hw-sensitivity}
\begin{figure}[t]
    \centering
    \includegraphics[width=\linewidth]{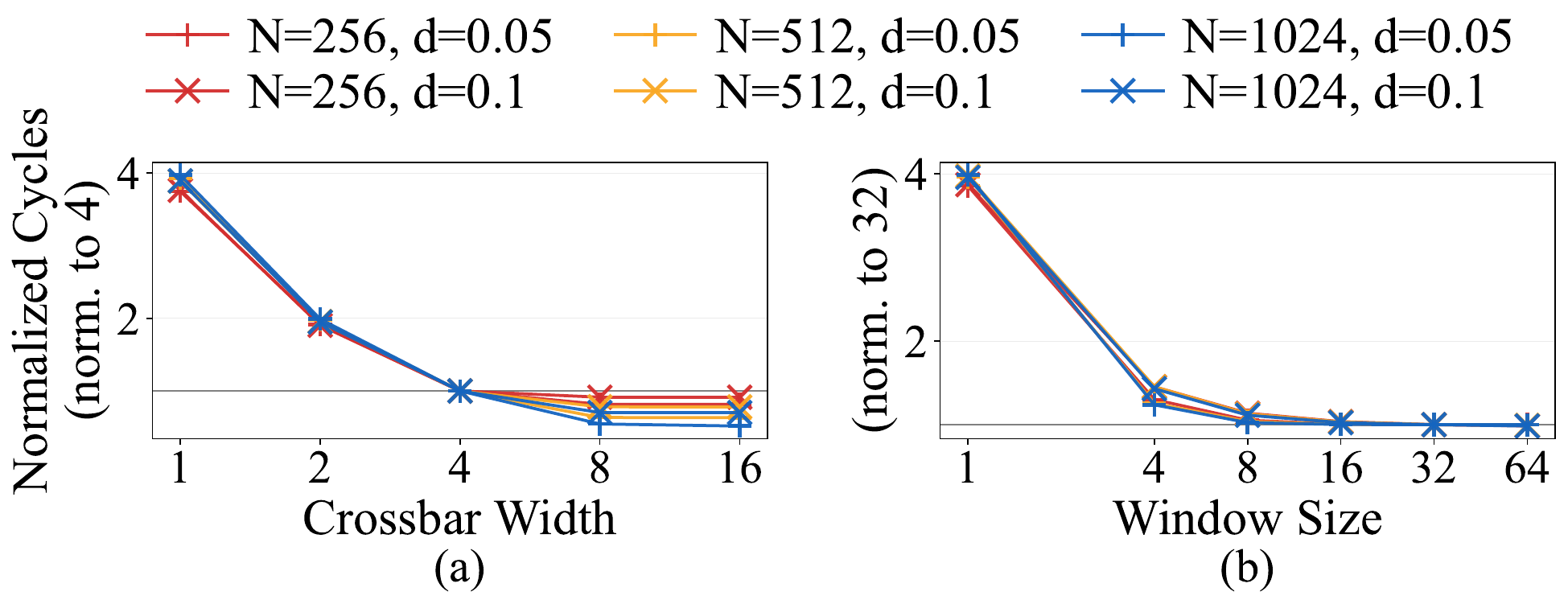}
    \vspace{-0.25in}
    \caption{Hardware-parameter sensitivity studies across three matrix sizes
        ($N\!=\!256, 512, 1024$) and two density levels ($d\!=\!0.05, 0.1$).
        \textbf{(a)}~Crossbar Width sweep, normalized to BRL$\!=\!4$.
        \textbf{(b)}~Window Size sweep, normalized to W$\!=\!32$.
        Color encodes matrix size, marker encodes density ($+$:
        $d\!=\!0.05$, $\times$: $d\!=\!0.1$).}
    \label{fig:sensitivity-combined}
    \vspace{-0.10in}
\end{figure}

To understand how hardware parameters shape SegFold's performance, we perform two sensitivity studies across three matrix sizes \{$256, 512, 1024$\} and two density levels \{$0.05, 0.1$\}: (i) varying the bandwidth of the vector multicast network that delivers $B$ rows, and (ii) varying the active-window size.

\smallskip\noindent\textbf{Vector Multicast Bandwidth.}
As shown in Fig.~\ref{fig:segfold-arch}(a), SegFold's global network routes $B$ rows from memory to PE rows through a vector multicast network. To observe sensitivity, we vary the network multicast width from 1 to 16 $B$ rows per cycle, keeping all other parameters fixed. As shown in Fig.~\ref{fig:sensitivity-combined}, performance improves noticeably from 1 to 4 rows per cycle across all matrix sizes and densities, but the marginal benefit beyond 4 rows per cycle quickly diminishes. At higher density ($d=0.1$), the sensitivity is more pronounced for larger matrices, as increased nonzero density creates more contention on the network. Based on this trend and the associated area/wiring cost, we choose a bandwidth of 4 $B$ rows per cycle.

\smallskip\noindent\textbf{Window Size of Active $B$ Rows.}
As discussed in $\S$\ref{sec:memory-controller}, the active window over $B$ rows controls how many $k$ values and $B$ rows can be reordered and scheduled in parallel. We sweep window sizes from 1 to 64, and report the normalized cycles in Fig.~\ref{fig:sensitivity-combined}. Performance improves substantially as the window grows up to 32, but shows little further gain beyond 32, consistently across both density levels. At higher density ($d=0.1$), the benefit of increasing window size is slightly more pronounced, as denser matrices expose more reordering opportunities. We therefore adopt a window size of 32 as our default configuration, as it offers a good trade-off between benefit and metadata/storage overhead.

\begin{figure}[t]
    \centering
    \includegraphics[width=0.9\linewidth]{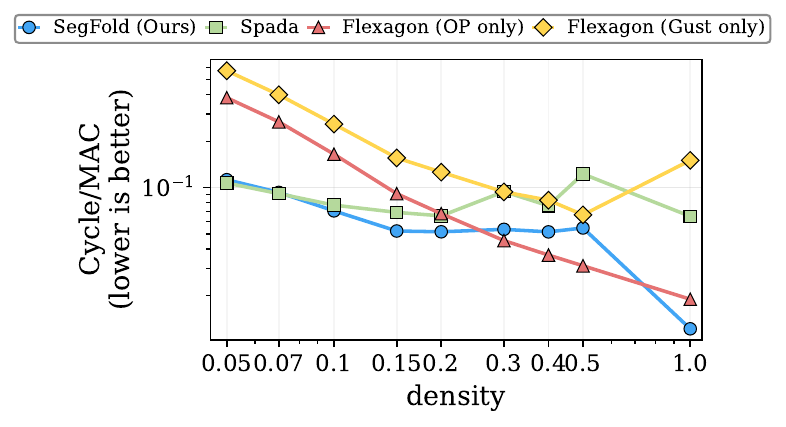}
    \vspace{-0.1in}
    \caption{Per-simulation efficiency (cycles per MAC, lower is better) on synthetic square matrices across densities, comparing Flexagon (OP and Gustavson), Spada, and SegFold. }
    \label{fig:density-sweep}
    \vspace{-0.25in}
\end{figure}

\subsubsection{Input Sensitivity}
\label{sec:input-sensitivity}
To understand how input characteristics affect SegFold's performance, we perform two sensitivity studies with the synthetic matrices: (i) varying the density of the two input matrices from highly sparse to entirely dense; and (ii) varying the relative sparsity of the two input matrices to create asymmetric sparsity patterns.

\smallskip\noindent\textbf{Density Sweep.}
We test synthetic matrices of sizes $256$ to $1024$ with densities ranging from $0.05$ to $1.0$. Figure~\ref{fig:density-sweep} reports the average cycles per MAC across the three accelerators. As density increases, SegFold's per-MAC efficiency stays roughly flat through mid densities and then drops sharply at the fully dense end, while Spada's degrades sharply once density exceeds $0.4$. This transition reflects bandwidth saturation in Spada's row-sequential 16-channel memory.
Flexagon's outer-product (OP) dataflow shows the biggest improvement with increasing density, as its static dataflow can better exploit reuse opportunities when more nonzeros are present, and the fixed control overhead is amortized. This also explains why SegFold's speedup over Flexagon is relatively large on the SuiteSparse matrices, which are generally sparser than the synthetic ones. Notably, in the fully dense case, SegFold outperforms all the baselines: its 2D array natively supports dense GEMM, and the dynamic-mapping overhead is diminished when the matrix is dense, since the $\mathcal{V}$ space mapping is essentially static---every position is occupied, leaving no shifts or skip decisions for the merge network to make, while baselines have the static overhead introduced by isolating the multiplication and reduction phases. For sparsity levels beyond the figure's range, SegFold's speedup over Spada increases. 

\begin{figure}[t]
    \centering
    \includegraphics[width=0.8\linewidth]{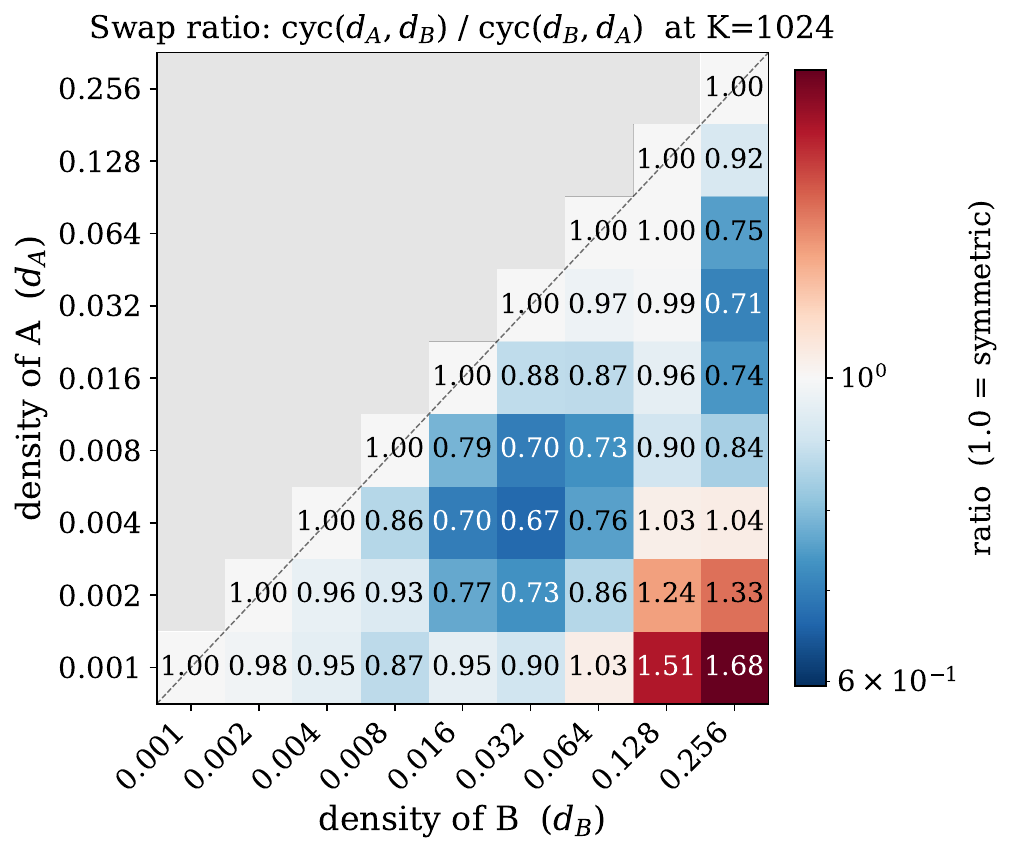}
    \vspace{-0.1in}
    \caption{Asymmetric-sparsity sensitivity at $K\!=\!1024$.}
    \label{fig:asymmetric-sparsity}
    \vspace{-0.3in}
\end{figure}

\smallskip\noindent\textbf{Asymmetric Sparsity.}
SegFold treats $A$ and $B$ asymmetrically, raising the question of whether sparsity differences between $A$ and $B$ influence whether $A\times B$ or $A^\top\times B^\top$ is faster. To understand this, we sweep $(d_A, d_B)$ pairs over synthetic matrices with size 1024 and report the swap ratio $\mathrm{cyc}(d_A, d_B)/\mathrm{cyc}(d_B, d_A)$ in Fig.~\ref{fig:asymmetric-sparsity}: ratios $<1$ (blue) favor placing the sparser matrix as operand $A$, while ratios $>1$ (red) favor placing the denser one there. Since the upper triangle is the reciprocal of the lower, we focus on the lower-right half ($d_A \le d_B$). Most of this region is blue: having the sparser matrix as operand $A$ is faster because \textsc{SelectA}'s fine-grained scheduling on $A$ leverages high sparsity better than the coarse-grained loader on $B$ can.

However, in the red corner where the disparity in density between $A$ and $B$ becomes very large, having the denser matrix as the first operand becomes faster. When $A$'s sparsity is extreme, a substantial fraction of $A$'s rows contain no nonzeros, yet each row still triggers a \textsc{SelectA} iteration with its associated scheduling and pipeline overhead; the fine-grained selection that benefits sparse $A$ in the common case now operates over many empty rows and produces no useful work. Swapping the operands places the denser matrix in operand $A$---every row is occupied, so no iteration is wasted---while the very sparse matrix moves to operand $B$, where the coarse-grained, demand-driven loader simply skips empty rows at no cost. The crossover from blue to red occurs once the density ratio $d_B/d_A$ grows large enough (in our experiments, around $32\times$ to $64\times$) that the per-row \textsc{SelectA} overhead saved by the swap exceeds the fine-grained sparsity benefit lost on $A$.

\subsection{RTL Results}
\label{sec:area}
Table~\ref{tab:area-power-summary} reports the post-synthesis RTL estimates of power and area for SegFold. We synthesize all modules using the ASAP7 7\,nm standard-cell library~\cite{clark2016asap7} at 1\,GHz. For the 2D mesh, we synthesize a single PE, switch, and FIFO buffer, and scale their results to a $16 \times 16$ grid. Each PE row includes a spad for overflow $C$ entries; these are synthesized per bank and scaled by 16 rows. The global memory controller is synthesized separately. At 7\,nm, the total design occupies 0.160\,mm$^2$ and consumes 385.7\,mW. To enable comparison with prior work synthesized at older technology nodes, we also provide estimated 28\,nm numbers by applying standard area ($\sim$8$\times$) and power ($\sim$5.5$\times$) scaling factors, with additional pipeline stages inserted to meet 1\,GHz timing at 28\,nm.


\begin{table}[t]
\caption{Power and area of SegFold. Left: ASAP7 7\,nm synthesis at 1\,GHz. Right: estimated 28\,nm scaling at 1\,GHz (with pipeline modifications).}
\footnotesize
\centering
\renewcommand{\arraystretch}{0.9}
\begin{tabular}{@{}l r r r r@{}}
\toprule
 & \multicolumn{2}{c}{\textbf{ASAP7 7\,nm}} & \multicolumn{2}{c}{\textbf{Est.\ 28\,nm}} \\
\cmidrule(lr){2-3} \cmidrule(lr){4-5}
\textbf{Component} & \textbf{Area} & \textbf{Power} & \textbf{Area} & \textbf{Power} \\
 & [$\mu$m$^2$] & [mW] & [$\mu$m$^2$] & [mW] \\
\midrule
PE ($\times$256)           &  26{,}304  &  59.1 & $\sim$231{,}400  & $\sim$342 \\
Switch ($\times$256)       &  10{,}967  &  27.1 &  $\sim$87{,}700  & $\sim$156 \\
FIFO Buffer ($\times$256)    & 105{,}600  & 263.4 & $\sim$887{,}000  & $\sim$1{,}574 \\
Scratchpad ($\times$16)     &  16{,}025  &  34.3 & $\sim$134{,}600  & $\sim$208 \\
Mem Controller ($\times$1) &   1{,}603  &   1.8 &  $\sim$13{,}470  & $\sim$11 \\
\midrule
\textbf{Total}  & \textbf{160{,}499} & \textbf{385.7} & $\sim$\textbf{1{,}354{,}200} & $\sim$\textbf{2{,}290} \\
 & (0.160\,mm$^2$) & & ($\sim$1.35\,mm$^2$) & \\
\bottomrule
\end{tabular}
\label{tab:area-power-summary}
\end{table}


\revc[C-Q2]{For comparison, Flexagon~\cite{flexagon} reports a total area (without cache) of 1.35\,mm$^2$ and power consumption of 856\,mW at 28\,nm. When scaled to the same 28\,nm technology node, SegFold's estimated area of $\sim$1.35\,mm$^2$ is comparable, while providing the additional hardware for dynamic scheduling (merge network, scratchpads) that enables the performance gains shown above. The estimated 28\,nm power of $\sim$2.3\,W assumes conservative vectorless 50\% switching activity; real sparse workloads exhibit lower activity factors, which would substantially reduce dynamic power.}

\section{Additional Related Work}

\smallskip\noindent\textbf{Reconfigurable Architectures.}
Reconfigurable architectures~\cite{plasticine,riptide,transmuter} are inherently programmable and can support different dataflows. Prior architectures have explored specialization for sparsity, such as SPU~\cite{spu} for sparse inner-product joins, and Capstan~\cite{capstan,stardust-capstan-compiler} for parallel indirect memory access in Gustavson.
To our knowledge, dynamic dataflow optimizations have not been explored on a reconfigurable architecture, and they would likely be costly given coarse-grain reconfigurable primitives.

\smallskip\noindent\textbf{Dynamic Prediction.}
Dynaflow~\cite{dynaflow} predicts the best dataflow for dataflow-flexible accelerators like those above, using ML techniques trained on dataset sparsity patterns.
SparseAdapt~\cite{sparse-adapt} is a runtime system for a flexible reconfigurable design called Transmuter~\cite{transmuter}, which can change microarchitectural policies depending on sparse data patterns.

\smallskip\noindent\textbf{Tile-Ordering Strategies.}
Inter-tile ordering strategies such as Hilbert or Z-order traversals---used in CUBE~\cite{cube} and Mosaic~\cite{mosaic}---and the sliding-window tiling in Spahet~\cite{Spahet} optimize the \emph{order} in which tiles are visited to improve cross-tile cache locality, but leave the scheduling of work \emph{within} a tile static. SegFold's dynamic scheduling is orthogonal: it operates at sub-tile granularity, reordering operations inside a tile based on runtime sparsity, and could therefore compose with any of these inter-tile ordering schemes.

\smallskip\noindent\textbf{Flexibility and Dynamism in Related Domains.}
EIE~\cite{eie} is a SpMV accelerator that mitigates load imbalance with large input queues. Cerebrus~\cite{cerebrus-spmv} is a SpMV accelerator that can support multiple dataflow patterns.  ZeNa~\cite{zena} is a sparse CNN accelerator that uses work-stealing queues to perform load balancing at a tile level. Graph processing accelerators face similar sparsity patterns and encounter the same load-balance and reuse challenges. PolyGraph~\cite{polygraph} uses offline preprocessing, which aims to improve both load balance and locality. HATS~\cite{hats} uses bounded depth-first search ordering to improve locality. AWB-GCN~\cite{awb-gcn} is a graph convolution accelerator, which dynamically load balances rows among PEs, and further splits pathologically long rows. These graph accelerators all exploit locality in settings where the output structure is known ahead of time; SegFold faces the additional challenge of discovering the output nonzero pattern on-the-fly during SpGEMM execution, requiring runtime locality decisions rather than offline preprocessing.

\smallskip\noindent\textbf{Comparison with Preprocessing-Based Approaches.}
Building on the Gustavson-style preprocessing designs introduced in $\S$\ref{sec:background}, we contrast SegFold's runtime approach with offline preprocessing in more detail. Gamma~\cite{gamma} and Zed~\cite{zed} use offline preprocessing to reorder rows of the stationary matrix, grouping rows with similar sparsity patterns to improve streaming-matrix reuse under a Gustavson dataflow. This preprocessing must be performed once per matrix and amortized across repeated SpGEMM invocations. In contrast, SegFold achieves similar reuse benefits---grouping $A$ columns with overlapping nonzero patterns via \textsc{SelectA}---entirely at runtime, with no preprocessing step. This makes SegFold particularly advantageous for workloads where the sparse matrices change frequently (e.g., dynamic graphs or iterative solvers), as the cost of preprocessing cannot be amortized. For static matrices used repeatedly, preprocessing-based approaches may offer complementary benefits, and combining offline reordering with SegFold's runtime scheduling is a promising direction for future work.

\section{Conclusion}

This work introduces Segment, a novel dynamic dataflow for SpGEMM that simultaneously optimizes data reuse and load balance, overcoming limitations of conventional static dataflows. Through the co-designed SegFold accelerator, we demonstrate that fine-grained dynamic scheduling and remapping of work across PEs can significantly improve performance and utilization across diverse sparsity patterns and matrix sizes. Our results---a $1.95\times$ geometric-mean speedup over the best dynamic baseline and $5.3\times$ over the best static baseline---highlight the value of incorporating dynamism into sparse-matrix accelerators, providing a new pathway for efficient execution of irregular workloads.


\bibliographystyle{IEEEtranS}
\bibliography{reference}

\begin{thebibliography}{10}
\providecommand{\url}[1]{#1}
\csname url@samestyle\endcsname
\providecommand{\newblock}{\relax}
\providecommand{\bibinfo}[2]{#2}
\providecommand{\BIBentrySTDinterwordspacing}{\spaceskip=0pt\relax}
\providecommand{\BIBentryALTinterwordstretchfactor}{4}
\providecommand{\BIBentryALTinterwordspacing}{\spaceskip=\fontdimen2\font plus
\BIBentryALTinterwordstretchfactor\fontdimen3\font minus
  \fontdimen4\font\relax}
\providecommand{\BIBforeignlanguage}[2]{{%
\expandafter\ifx\csname l@#1\endcsname\relax
\typeout{** WARNING: IEEEtranS.bst: No hyphenation pattern has been}%
\typeout{** loaded for the language `#1'. Using the pattern for}%
\typeout{** the default language instead.}%
\else
\language=\csname l@#1\endcsname
\fi
#2}}
\providecommand{\BIBdecl}{\relax}
\BIBdecl

\bibitem{dcsr}
A.~Buluç and J.~R. Gilbert, ``On the representation and multiplication of
  hypersparse matrices,'' in \emph{2008 IEEE International Symposium on
  Parallel and Distributed Processing (IPDPS)}, 2008, pp. 1--11.

\bibitem{clark2016asap7}
L.~T. Clark, V.~Vashishtha, L.~Shifren, A.~Gujja, S.~Sinha, B.~Cline,
  C.~Ramamurthy, and G.~Yeric, ``{ASAP7}: A 7-nm {FinFET} predictive process
  design kit,'' \emph{Microelectronics Journal}, vol.~53, pp. 105--115, 2016.

\bibitem{gpu}
N.~Corp, ``{NVIDIA GPU programming guide, v2.2.1, November, 2004}.''

\bibitem{polygraph}
V.~Dadu, S.~Liu, and T.~Nowatzki, ``{PolyGraph}: Exposing the value of
  flexibility for graph processing accelerators,'' in \emph{2021 ACM/IEEE 48th
  Annual International Symposium on Computer Architecture (ISCA)}, 2021, pp.
  595--608.

\bibitem{spu}
\BIBentryALTinterwordspacing
V.~Dadu, J.~Weng, S.~Liu, and T.~Nowatzki, ``Towards general purpose
  acceleration by exploiting common data-dependence forms,'' ser. MICRO
  '52.\hskip 1em plus 0.5em minus 0.4em\relax New York, NY, USA: ACM, 2019, pp.
  924--939. [Online]. Available:
  \url{http://doi.acm.org/10.1145/3352460.3358276}
\BIBentrySTDinterwordspacing

\bibitem{zed}
\BIBentryALTinterwordspacing
P.~Dangi, Z.~Bai, R.~Juneja, D.~Wijerathne, and T.~Mitra, ``{ZeD}: A
  generalized accelerator for variably sparse matrix computations in {ML},'' in
  \emph{Proceedings of the 2024 International Conference on Parallel
  Architectures and Compilation Techniques}, ser. PACT '24.\hskip 1em plus
  0.5em minus 0.4em\relax New York, NY, USA: Association for Computing
  Machinery, 2024, p. 246–257. [Online]. Available:
  \url{https://doi.org/10.1145/3656019.3689905}
\BIBentrySTDinterwordspacing

\bibitem{daveSurvey}
S.~Dave, R.~Baghdadi, T.~Nowatzki, S.~Avancha, A.~Shrivastava, and B.~Li,
  ``Hardware acceleration of sparse and irregular tensor computations of {ML}
  models: A survey and insights,'' \emph{Proceedings of the IEEE}, vol. 109,
  no.~10, pp. 1706--1752, 2021.

\bibitem{suitesparse}
T.~A. Davis and Y.~Hu, ``The university of florida sparse matrix collection,''
  \emph{ACM Transactions on Mathematical Software}, vol.~38, no.~1, pp.
  1:1--1:25, 2011.

\bibitem{tpu2}
J.~Dean, ``Recent advances in artificial intelligence via machine learning and
  the implications for computer system design,'' in \emph{2017 IEEE Hot Chips
  29 Symposium}, 2017.

\bibitem{spgemm-survey}
\BIBentryALTinterwordspacing
J.~Gao, W.~Ji, F.~Chang, S.~Han, B.~Wei, Z.~Liu, and Y.~Wang, ``A systematic
  survey of general sparse {Matrix-Matrix} multiplication,'' \emph{ACM Comput.
  Surv.}, vol.~55, no.~12, Mar. 2023. [Online]. Available:
  \url{https://doi.org/10.1145/3571157}
\BIBentrySTDinterwordspacing

\bibitem{stellar}
\BIBentryALTinterwordspacing
H.~N. Genc, H.~Kim, P.~Ganesh, and Y.~S. Shao, ``{Stellar}: An automated design
  framework for dense and sparse spatial accelerators,'' in \emph{Proceedings
  of the 2024 57th IEEE/ACM International Symposium on Microarchitecture}, ser.
  MICRO '24.\hskip 1em plus 0.5em minus 0.4em\relax IEEE Press, 2024, p.
  409–422. [Online]. Available:
  \url{https://doi.org/10.1109/MICRO61859.2024.00038}
\BIBentrySTDinterwordspacing

\bibitem{awb-gcn}
T.~Geng, A.~Li, R.~Shi, C.~Wu, T.~Wang, Y.~Li, P.~Haghi, A.~Tumeo, S.~Che,
  S.~Reinhardt, and M.~C. Herbordt, ``{AWB-GCN}: A graph convolutional network
  accelerator with runtime workload rebalancing,'' in \emph{2020 53rd Annual
  IEEE/ACM International Symposium on Microarchitecture (MICRO)}, 2020, pp.
  922--936.

\bibitem{riptide}
G.~Gobieski, S.~Ghosh, M.~Heule, T.~Mowry, T.~Nowatzki, N.~Beckmann, and
  B.~Lucia, ``{RipTide}: A programmable, energy-minimal dataflow compiler and
  architecture,'' in \emph{2022 55th IEEE/ACM International Symposium on
  Microarchitecture (MICRO)}, 2022, pp. 546--564.

\bibitem{sparten}
\BIBentryALTinterwordspacing
A.~Gondimalla, N.~Chesnut, M.~Thottethodi, and T.~N. Vijaykumar, ``{SparTen}: A
  sparse tensor accelerator for convolutional neural networks,'' in
  \emph{Proceedings of the 52nd Annual IEEE/ACM International Symposium on
  Microarchitecture}, ser. MICRO '52.\hskip 1em plus 0.5em minus 0.4em\relax
  New York, NY, USA: Association for Computing Machinery, 2019, p. 151–165.
  [Online]. Available: \url{https://doi.org/10.1145/3352460.3358291}
\BIBentrySTDinterwordspacing

\bibitem{eie}
\BIBentryALTinterwordspacing
S.~Han, X.~Liu, H.~Mao, J.~Pu, A.~Pedram, M.~A. Horowitz, and W.~J. Dally,
  ``Eie: efficient inference engine on compressed deep neural network,'' in
  \emph{Proceedings of the 43rd International Symposium on Computer
  Architecture}, ser. ISCA '16.\hskip 1em plus 0.5em minus 0.4em\relax IEEE
  Press, 2016, p. 243–254. [Online]. Available:
  \url{https://doi.org/10.1109/ISCA.2016.30}
\BIBentrySTDinterwordspacing

\bibitem{he2020sparse}
X.~He, S.~Pal, A.~Amarnath, S.~Feng, D.-H. Park, A.~Rovinski, H.~Ye, Y.~Chen,
  R.~Dreslinski, and T.~Mudge, ``{Sparse-TPU}: Adapting systolic arrays for
  sparse matrices,'' in \emph{Proceedings of the 34th ACM international
  conference on supercomputing}, 2020, pp. 1--12.

\bibitem{extensor}
\BIBentryALTinterwordspacing
K.~Hegde, H.~Asghari-Moghaddam, M.~Pellauer, N.~Crago, A.~Jaleel, E.~Solomonik,
  J.~Emer, and C.~W. Fletcher, ``{ExTensor}: An accelerator for sparse tensor
  algebra,'' in \emph{Proceedings of the 52nd Annual IEEE/ACM International
  Symposium on Microarchitecture}, ser. MICRO-52.\hskip 1em plus 0.5em minus
  0.4em\relax New York, NY, USA: Association for Computing Machinery, 2019, p.
  319–333. [Online]. Available: \url{https://doi.org/10.1145/3352460.3358275}
\BIBentrySTDinterwordspacing

\bibitem{stardust-capstan-compiler}
\BIBentryALTinterwordspacing
O.~Hsu, A.~Rucker, T.~Zhao, V.~Desai, K.~Olukotun, and F.~Kjolstad,
  ``{Stardust}: Compiling sparse tensor algebra to a reconfigurable dataflow
  architecture,'' in \emph{Proceedings of the 23rd ACM/IEEE International
  Symposium on Code Generation and Optimization}, ser. CGO '25.\hskip 1em plus
  0.5em minus 0.4em\relax New York, NY, USA: Association for Computing
  Machinery, 2025, p. 628–643. [Online]. Available:
  \url{https://doi.org/10.1145/3696443.3708918}
\BIBentrySTDinterwordspacing

\bibitem{Spahet}
\BIBentryALTinterwordspacing
H.~Huang, P.~Yao, Z.~An, Y.~Sun, A.~Hu, P.~Xu, L.~Zheng, X.~Liao, and H.~Jin,
  ``{SpaHet}: A software/hardware co-design for accelerating
  heterogeneous-sparsity based sparse matrix multiplication,'' in
  \emph{Proceedings of the 61st ACM/IEEE Design Automation Conference}, ser.
  DAC '24.\hskip 1em plus 0.5em minus 0.4em\relax New York, NY, USA:
  Association for Computing Machinery, 2024. [Online]. Available:
  \url{https://doi.org/10.1145/3649329.3655944}
\BIBentrySTDinterwordspacing

\bibitem{cerebrus-spmv}
\BIBentryALTinterwordspacing
S.~Hwang, D.~Baek, J.~Park, and J.~Huh, ``{Cerberus}: Triple mode acceleration
  of sparse matrix and vector multiplication,'' \emph{ACM Trans. Archit. Code
  Optim.}, vol.~21, no.~2, May 2024. [Online]. Available:
  \url{https://doi.org/10.1145/3653020}
\BIBentrySTDinterwordspacing

\bibitem{tpu}
\BIBentryALTinterwordspacing
N.~P. Jouppi, C.~Young, N.~Patil, D.~Patterson, G.~Agrawal, R.~Bajwa, S.~Bates,
  S.~Bhatia, N.~Boden, A.~Borchers, R.~Boyle, P.-l. Cantin, C.~Chao, C.~Clark,
  J.~Coriell, M.~Daley, M.~Dau, J.~Dean, B.~Gelb, T.~V. Ghaemmaghami,
  R.~Gottipati, W.~Gulland, R.~Hagmann, C.~R. Ho, D.~Hogberg, J.~Hu, R.~Hundt,
  D.~Hurt, J.~Ibarz, A.~Jaffey, A.~Jaworski, A.~Kaplan, H.~Khaitan,
  D.~Killebrew, A.~Koch, N.~Kumar, S.~Lacy, J.~Laudon, J.~Law, D.~Le, C.~Leary,
  Z.~Liu, K.~Lucke, A.~Lundin, G.~MacKean, A.~Maggiore, M.~Mahony, K.~Miller,
  R.~Nagarajan, R.~Narayanaswami, R.~Ni, K.~Nix, T.~Norrie, M.~Omernick,
  N.~Penukonda, A.~Phelps, J.~Ross, M.~Ross, A.~Salek, E.~Samadiani, C.~Severn,
  G.~Sizikov, M.~Snelham, J.~Souter, D.~Steinberg, A.~Swing, M.~Tan,
  G.~Thorson, B.~Tian, H.~Toma, E.~Tuttle, V.~Vasudevan, R.~Walter, W.~Wang,
  E.~Wilcox, and D.~H. Yoon, ``In-datacenter performance analysis of a tensor
  processing unit,'' in \emph{Proceedings of the 44th Annual International
  Symposium on Computer Architecture}, ser. ISCA '17.\hskip 1em plus 0.5em
  minus 0.4em\relax New York, NY, USA: ACM, 2017, pp. 1--12. [Online].
  Available: \url{http://doi.acm.org/10.1145/3079856.3080246}
\BIBentrySTDinterwordspacing

\bibitem{zena}
D.~Kim, J.~Ahn, and S.~Yoo, ``{Zena}: Zero-aware neural network accelerator,''
  \emph{IEEE Design \& Test}, vol.~35, no.~1, pp. 39--46, 2017.

\bibitem{koul2024onyx}
K.~Koul, M.~Strange, J.~Melchert, A.~Carsello, Y.~Mei, O.~Hsu, T.~Kong, P.-H.
  Chen, H.~Ke, K.~Zhang \emph{et~al.}, ``{Onyx}: A programmable accelerator for
  sparse tensor algebra,'' in \emph{2024 IEEE Hot Chips 36 Symposium
  (HCS)}.\hskip 1em plus 0.5em minus 0.4em\relax IEEE Computer Society, 2024,
  pp. 1--91.

\bibitem{spada}
\BIBentryALTinterwordspacing
Z.~Li, J.~Li, T.~Chen, D.~Niu, H.~Zheng, Y.~Xie, and M.~Gao, ``{Spada}:
  Accelerating sparse matrix multiplication with adaptive dataflow,'' in
  \emph{Proceedings of the 28th ACM International Conference on Architectural
  Support for Programming Languages and Operating Systems, Volume 2}, ser.
  ASPLOS 2023.\hskip 1em plus 0.5em minus 0.4em\relax New York, NY, USA:
  Association for Computing Machinery, 2023, p. 747–761. [Online]. Available:
  \url{https://doi.org/10.1145/3575693.3575706}
\BIBentrySTDinterwordspacing

\bibitem{ramulator2}
\BIBentryALTinterwordspacing
H.~Luo, Y.~C. Tuğrul, F.~N. Bostancı, A.~Olgun, A.~G. Yağlıkçı, and
  O.~Mutlu, ``{Ramulator} 2.0: A modern, modular, and extensible {DRAM}
  simulator,'' 2023. [Online]. Available:
  \url{https://arxiv.org/abs/2308.11030}
\BIBentrySTDinterwordspacing

\bibitem{sparm}
S.~Luo, B.~Wang, Y.~Shi, X.~Zhang, Q.~Xue, and S.~Ma, ``{Sparm}: A sparse
  matrix multiplication accelerator supporting multiple dataflows,'' in
  \emph{2024 IEEE 35th International Conference on Application-specific
  Systems, Architectures and Processors (ASAP)}, 2024, pp. 122--130.

\bibitem{mosaic}
\BIBentryALTinterwordspacing
S.~Maass, C.~Min, S.~Kashyap, W.~Kang, M.~Kumar, and T.~Kim, ``{Mosaic}:
  Processing a trillion-edge graph on a single machine,'' in \emph{Proceedings
  of the Twelfth European Conference on Computer Systems}, ser. EuroSys
  '17.\hskip 1em plus 0.5em minus 0.4em\relax New York, NY, USA: Association
  for Computing Machinery, 2017, p. 527–543. [Online]. Available:
  \url{https://doi.org/10.1145/3064176.3064191}
\BIBentrySTDinterwordspacing

\bibitem{flexagon}
\BIBentryALTinterwordspacing
F.~Mu\~{n}oz Mart\'{\i}nez, R.~Garg, M.~Pellauer, J.~L. Abell\'{a}n, M.~E.
  Acacio, and T.~Krishna, ``{Flexagon}: A multi-dataflow sparse-sparse matrix
  multiplication accelerator for efficient {DNN} processing,'' in
  \emph{Proceedings of the 28th ACM International Conference on Architectural
  Support for Programming Languages and Operating Systems, Volume 3}, ser.
  ASPLOS 2023.\hskip 1em plus 0.5em minus 0.4em\relax New York, NY, USA:
  Association for Computing Machinery, 2023, p. 252–265. [Online]. Available:
  \url{https://doi.org/10.1145/3582016.3582069}
\BIBentrySTDinterwordspacing

\bibitem{hats}
A.~Mukkara, N.~Beckmann, M.~Abeydeera, X.~Ma, and D.~Sanchez, ``Exploiting
  locality in graph analytics through hardware-accelerated traversal
  scheduling,'' in \emph{MICRO}.\hskip 1em plus 0.5em minus 0.4em\relax IEEE,
  2018, pp. 1--14.

\bibitem{stonne}
F.~Muñoz-Martínez, J.~L. Abellán, M.~E. Acacio, and T.~Krishna, ``{STONNE}:
  Enabling cycle-level microarchitectural simulation for {DNN} inference
  accelerators,'' in \emph{2021 IEEE International Symposium on Workload
  Characterization (IISWC)}, 2021, pp. 201--213.

\bibitem{teaal}
\BIBentryALTinterwordspacing
N.~Nayak, T.~O. Odemuyiwa, S.~Ugare, C.~Fletcher, M.~Pellauer, and J.~Emer,
  ``{TeAAL}: A declarative framework for modeling sparse tensor accelerators,''
  in \emph{Proceedings of the 56th Annual IEEE/ACM International Symposium on
  Microarchitecture}, ser. MICRO '23.\hskip 1em plus 0.5em minus 0.4em\relax
  New York, NY, USA: Association for Computing Machinery, 2023, p. 1255–1270.
  [Online]. Available: \url{https://doi.org/10.1145/3613424.3623791}
\BIBentrySTDinterwordspacing

\bibitem{svm-accel}
E.~Nurvitadhi, A.~Mishra, and D.~Marr, ``A sparse matrix vector multiply
  accelerator for support vector machine,'' in \emph{2015 International
  Conference on Compilers, Architecture and Synthesis for Embedded Systems
  (CASES)}, Oct 2015, pp. 109--116.

\bibitem{outerspace}
S.~Pal, J.~Beaumont, D.~Park, A.~Amarnath, S.~Feng, C.~Chakrabarti, H.~Kim,
  D.~Blaauw, T.~Mudge, and R.~Dreslinski, ``{OuterSPACE}: An outer product
  based sparse matrix multiplication accelerator,'' in \emph{2018 IEEE
  International Symposium on High Performance Computer Architecture (HPCA)},
  Feb 2018, pp. 724--736.

\bibitem{sparse-adapt}
\BIBentryALTinterwordspacing
S.~Pal, A.~Amarnath, S.~Feng, M.~O'Boyle, R.~Dreslinski, and C.~Dubach,
  ``{SparseAdapt}: Runtime control for sparse linear algebra on a
  reconfigurable accelerator,'' in \emph{MICRO-54: 54th Annual IEEE/ACM
  International Symposium on Microarchitecture}, ser. MICRO '21.\hskip 1em plus
  0.5em minus 0.4em\relax New York, NY, USA: Association for Computing
  Machinery, 2021, p. 1005–1021. [Online]. Available:
  \url{https://doi.org/10.1145/3466752.3480134}
\BIBentrySTDinterwordspacing

\bibitem{transmuter}
\BIBentryALTinterwordspacing
S.~Pal, S.~Feng, D.-h. Park, S.~Kim, A.~Amarnath, C.-S. Yang, X.~He,
  J.~Beaumont, K.~May, Y.~Xiong, K.~Kaszyk, J.~M. Morton, J.~Sun, M.~O'Boyle,
  M.~Cole, C.~Chakrabarti, D.~Blaauw, H.-S. Kim, T.~Mudge, and R.~Dreslinski,
  ``{Transmuter}: Bridging the efficiency gap using memory and dataflow
  reconfiguration,'' in \emph{Proceedings of the ACM International Conference
  on Parallel Architectures and Compilation Techniques}, ser. PACT '20.\hskip
  1em plus 0.5em minus 0.4em\relax New York, NY, USA: Association for Computing
  Machinery, 2020, p. 175–190. [Online]. Available:
  \url{https://doi.org/10.1145/3410463.3414627}
\BIBentrySTDinterwordspacing

\bibitem{scnn}
\BIBentryALTinterwordspacing
A.~Parashar, M.~Rhu, A.~Mukkara, A.~Puglielli, R.~Venkatesan, B.~Khailany,
  J.~Emer, S.~W. Keckler, and W.~J. Dally, ``{SCNN}: An accelerator for
  compressed-sparse convolutional neural networks,'' in \emph{Proceedings of
  the 44th Annual International Symposium on Computer Architecture}, ser. ISCA
  '17.\hskip 1em plus 0.5em minus 0.4em\relax New York, NY, USA: ACM, 2017, pp.
  27--40. [Online]. Available: \url{http://doi.acm.org/10.1145/3079856.3080254}
\BIBentrySTDinterwordspacing

\bibitem{plasticine}
\BIBentryALTinterwordspacing
R.~Prabhakar, Y.~Zhang, D.~Koeplinger, M.~Feldman, T.~Zhao, S.~Hadjis,
  A.~Pedram, C.~Kozyrakis, and K.~Olukotun, ``Plasticine: A reconfigurable
  architecture for parallel paterns,'' ser. ISCA '17.\hskip 1em plus 0.5em
  minus 0.4em\relax New York, NY, USA: ACM, 2017, pp. 389--402. [Online].
  Available: \url{http://doi.acm.org/10.1145/3079856.3080256}
\BIBentrySTDinterwordspacing

\bibitem{multi-compression}
E.~Qin, G.~Jeong, W.~Won, S.-C. Kao, H.~Kwon, S.~Srinivasan, D.~Das, G.~E.
  Moon, S.~Rajamanickam, and T.~Krishna, ``Extending sparse tensor accelerators
  to support multiple compression formats,'' in \emph{2021 IEEE International
  Parallel and Distributed Processing Symposium (IPDPS)}.\hskip 1em plus 0.5em
  minus 0.4em\relax IEEE, 2021, pp. 1014--1024.

\bibitem{sigma}
E.~Qin, A.~Samajdar, H.~Kwon, V.~Nadella, S.~Srinivasan, D.~Das, B.~Kaul, and
  T.~Krishna, ``{SIGMA}: A sparse and irregular {GEMM} accelerator with
  flexible interconnects for {DNN} training,'' in \emph{2020 IEEE International
  Symposium on High Performance Computer Architecture (HPCA)}, 2020, pp.
  58--70.

\bibitem{capstan}
A.~Rucker, M.~Vilim, T.~Zhao, Y.~Zhang, R.~Prabhakar, and K.~Olukotun,
  ``{Capstan}: A vector {RDA} for sparsity,'' 2021.

\bibitem{song2022serpens}
L.~Song, Y.~Chi, L.~Guo, and J.~Cong, ``{Serpens}: A high bandwidth memory
  based accelerator for general-purpose sparse matrix-vector multiplication,''
  in \emph{Proceedings of the 59th ACM/IEEE design automation conference},
  2022, pp. 211--216.

\bibitem{song2022sextans}
L.~Song, Y.~Chi, A.~Sohrabizadeh, Y.-k. Choi, J.~Lau, and J.~Cong, ``{Sextans}:
  A streaming accelerator for general-purpose sparse-matrix dense-matrix
  multiplication,'' in \emph{Proceedings of the 2022 ACM/SIGDA International
  Symposium on Field-Programmable Gate Arrays}, 2022, pp. 65--77.

\bibitem{matraptor}
N.~Srivastava, H.~Jin, J.~Liu, D.~Albonesi, and Z.~Zhang, ``{MatRaptor}: A
  sparse-sparse matrix multiplication accelerator based on row-wise product,''
  in \emph{2020 53rd Annual IEEE/ACM International Symposium on
  Microarchitecture (MICRO)}, 2020, pp. 766--780.

\bibitem{tensaurus}
N.~Srivastava, H.~Jin, S.~Smith, H.~Rong, D.~Albonesi, and Z.~Zhang,
  ``{Tensaurus}: A versatile accelerator for mixed sparse-dense tensor
  computations,'' in \emph{2020 IEEE International Symposium on High
  Performance Computer Architecture (HPCA)}.\hskip 1em plus 0.5em minus
  0.4em\relax IEEE, 2020, pp. 689--702.

\bibitem{sparGD}
\BIBentryALTinterwordspacing
B.~Wang, S.~Ma, S.~Luo, L.~Wu, J.~Zhang, C.~Zhang, and T.~Li, ``{SparGD}: A
  sparse {GEMM} accelerator with dynamic dataflow,'' \emph{ACM Trans. Des.
  Autom. Electron. Syst.}, vol.~29, no.~2, Jan. 2024. [Online]. Available:
  \url{https://doi.org/10.1145/3634703}
\BIBentrySTDinterwordspacing

\bibitem{spmard}
\BIBentryALTinterwordspacing
B.~Wang, S.~Ma, Y.~Zhao, S.~Luo, L.~Wu, J.~Zhang, D.~Li, T.~Li, and Z.~Chen,
  ``{SpMARD}: A sparse-sparse matrix multiplication accelerator with
  reconfigurable dataflow for {DNN} workloads,'' \emph{ACM Trans. Archit. Code
  Optim.}, Aug. 2025, just Accepted. [Online]. Available:
  \url{https://doi.org/10.1145/3747847}
\BIBentrySTDinterwordspacing

\bibitem{sparseloop}
\BIBentryALTinterwordspacing
Y.~N. Wu, P.-A. Tsai, A.~Parashar, V.~Sze, and J.~S. Emer, ``{Sparseloop}: An
  analytical approach to sparse tensor accelerator modeling,'' in
  \emph{Proceedings of the 55th Annual IEEE/ACM International Symposium on
  Microarchitecture}, ser. MICRO '22.\hskip 1em plus 0.5em minus 0.4em\relax
  IEEE Press, 2023, p. 1377–1395. [Online]. Available:
  \url{https://doi.org/10.1109/MICRO56248.2022.00096}
\BIBentrySTDinterwordspacing

\bibitem{dynaflow}
S.~Yadav and B.~Asgari, ``{DynaFlow}: An {ML} framework for dynamic dataflow
  selection in {SpGEMM} accelerators,'' \emph{IEEE Computer Architecture
  Letters}, vol.~24, no.~1, pp. 189--192, 2025.

\bibitem{trapezoid}
Y.~Yang, J.~S. Emer, and D.~Sanchez, ``Trapezoid: A versatile accelerator for
  dense and sparse matrix multiplications,'' in \emph{2024 ACM/IEEE 51st Annual
  International Symposium on Computer Architecture (ISCA)}, 2024, pp. 931--945.

\bibitem{scalagraph}
P.~Yao, L.~Zheng, Y.~Huang, Q.~Wang, C.~Gui, Z.~Zeng, X.~Liao, H.~Jin, and
  J.~Xue, ``{ScalaGraph}: A scalable accelerator for massively parallel graph
  processing,'' in \emph{2022 IEEE International Symposium on High-Performance
  Computer Architecture (HPCA)}, 2022, pp. 199--212.

\bibitem{gamma}
\BIBentryALTinterwordspacing
G.~Zhang, N.~Attaluri, J.~S. Emer, and D.~Sanchez, ``{Gamma}: Leveraging
  {Gustavson's} algorithm to accelerate sparse matrix multiplication,'' ser.
  ASPLOS 2021.\hskip 1em plus 0.5em minus 0.4em\relax New York, NY, USA:
  Association for Computing Machinery, 2021. [Online]. Available:
  \url{https://doi.org/10.1145/3445814.3446702}
\BIBentrySTDinterwordspacing

\bibitem{cube}
M.~Zhang, Y.~Wu, K.~Chen, X.~Qian, X.~Li, and W.~Zheng, ``Exploring the hidden
  dimension in graph processing,'' in \emph{Proceedings of the 12th USENIX
  Conference on Operating Systems Design and Implementation}, ser.
  OSDI'16.\hskip 1em plus 0.5em minus 0.4em\relax USA: USENIX Association,
  2016, p. 285–300.

\bibitem{cambricon-x}
S.~Zhang, Z.~Du, L.~Zhang, H.~Lan, S.~Liu, L.~Li, Q.~Guo, T.~Chen, and Y.~Chen,
  ``{Cambricon-X}: An accelerator for sparse neural networks,'' in \emph{2016
  49th Annual IEEE/ACM International Symposium on Microarchitecture (MICRO)},
  2016, pp. 1--12.

\bibitem{sparch}
Z.~Zhang, H.~Wang, S.~Han, and W.~J. Dally, ``{Sparch}: Efficient architecture
  for sparse matrix multiplication,'' in \emph{2020 IEEE International
  Symposium on High Performance Computer Architecture (HPCA)}.\hskip 1em plus
  0.5em minus 0.4em\relax IEEE, 2020, pp. 261--274.

\bibitem{feasta}
K.~Zhong, Z.~Zhu, G.~Dai, H.~Wang, X.~Yang, H.~Zhang, J.~Si, Q.~Mao, S.~Zeng,
  K.~Hong \emph{et~al.}, ``{Feasta}: A flexible and efficient accelerator for
  sparse tensor algebra in machine learning,'' in \emph{Proceedings of the 29th
  ACM International Conference on Architectural Support for Programming
  Languages and Operating Systems, Volume 3}, 2024, pp. 349--366.

\bibitem{cambricon-s}
X.~Zhou, Z.~Du, Q.~Guo, S.~Liu, C.~Liu, C.~Wang, X.~Zhou, L.~Li, T.~Chen, and
  Y.~Chen, ``{Cambricon-S}: Addressing irregularity in sparse neural networks
  through a cooperative software/hardware approach,'' in \emph{2018 51st Annual
  IEEE/ACM International Symposium on Microarchitecture (MICRO)}, 2018, pp.
  15--28.

\end{thebibliography}

\newpage
\appendix

\subsection*{Abstract}

We release the complete SegFold artifact, consisting of a cycle-accurate C++ simulator, real-world and synthetic benchmark suites, RTL source with synthesis reports, and automated scripts that regenerate every figure and table in the paper.
Concretely, the artifact contains: (i)~\texttt{csegfold}, a simulator that faithfully models SegFold's microarchitecture and dynamic dataflow, (ii)~a curated set of sparse matrices drawn from the SuiteSparse collection, (iii)~SystemVerilog RTL and corresponding area/power/timing reports for all hardware modules, and (iv)~end-to-end automation that builds the simulator, runs all experiments, and produces publication-ready plots.

\subsection*{Artifact Check-List (Meta-Information)}

{\small
\begin{itemize}[nosep,leftmargin=*]
  \item \textbf{Algorithm:} Segment dataflow for SpGEMM with fine-grained dynamic scheduling and work remapping.
  \item \textbf{Program:} \texttt{csegfold} --- cycle-accurate C++ simulator.
  \item \textbf{Compilation:} CMake $\geq$ 3.15, GCC 10+ (C++20 required). Ramulator2 is pulled automatically during the build via CMake FetchContent.
  \item \textbf{Binary:} \texttt{csegfold/build/csegfold}, compiled from source.
  \item \textbf{Data set:} $\sim$20 SuiteSparse matrices ($\sim$50\,MB, auto-downloaded), covering the 15-matrix baseline suite plus the additional matrices used in ablation studies; synthetic matrices are created on-the-fly.
  \item \textbf{Hardware:} Commodity server --- $\geq$4 CPU cores and $\geq$64\,GB RAM (16+ cores and 256\,GB RAM recommended for full parallelism).
  \item \textbf{Run-time environment:} Linux (tested on Ubuntu 22.04+), Python $\geq$ 3.8. A Docker image is also available.
  \item \textbf{Metrics:} Simulated cycle counts and speedup relative to prior accelerators.
  \item \textbf{Output:} Per-experiment CSV files and PDF/PNG plots corresponding to Figures~8--12.
  \item \textbf{Experiments:} 209 individual simulation runs, completing in $\sim$2 hours on a 16-core machine.
  \item \textbf{How much disk space required?} $\sim$2\,GB (source, benchmarks, and generated outputs).
  \item \textbf{How much time is needed to prepare the workflow?} $\sim$5 minutes for building the simulator and downloading matrices.
  \item \textbf{How much time is needed to complete experiments?} $\sim$2 hours at 16 cores; runtime scales roughly linearly with core count.
  \item \textbf{Publicly available?} Yes.
  \item \textbf{Code licenses?} MIT License.
  \item \textbf{Archived?} Yes (GitHub + Zenodo) DOI: 10.5281/zenodo.19453259.
\end{itemize}
}

\subsection*{Description}

\subsubsection*{How to Access}

The artifact is hosted on GitHub: \\
\url{https://github.com/PolyArch/SegFold-AE}

and also Zenodo: \\
\url{https://doi.org/10.5281/zenodo.19453259}

\subsubsection*{Hardware Dependencies}

\begin{itemize}[nosep,leftmargin=*]
  \item CPU: 4 cores minimum, 16+ cores recommended.
  \item RAM: 64\,GB minimum, 256\,GB recommended. Memory-intensive experiments (breakdown and mapping ablation) may consume up to 50\,GB per process.
  \item Disk: 2--5\,GB.
\end{itemize}

\subsubsection*{Software Dependencies}

\begin{itemize}[nosep,leftmargin=*]
  \item OS: Ubuntu 22.04 or later (other Linux distributions are expected to work).
  \item Toolchain: GCC 10+ (C++20), CMake $\geq$ 3.15.
  \item Python $\geq$ 3.8 with numpy, scipy, matplotlib, pandas, and pyyaml.
  \item Docker (optional, for a self-contained environment).
\end{itemize}

\subsubsection*{Data Sets}

The experiments use 20 sparse matrices from the SuiteSparse Matrix Collection (\url{https://sparse.tamu.edu/}), fetched automatically by a provided download script. Synthetic matrices for sensitivity studies are generated at runtime.

\subsection*{Installation}

\smallskip\noindent\textbf{Native build.}
\begin{enumerate}[nosep,leftmargin=*]
  \item Build the simulator and run a smoke test:
\begin{verbatim}
./scripts/setup.sh
\end{verbatim}
  \item Download benchmark matrices:
\begin{verbatim}
python3 scripts/download_matrices.py
\end{verbatim}
\end{enumerate}

\smallskip\noindent\textbf{Docker (alternative).} A pre-configured container is also available:
\begin{verbatim}
docker compose build
docker compose run artifact \
  ./scripts/run_all.sh
\end{verbatim}

\subsection*{Experiment Workflow}

A single command reproduces every experiment:
\begin{verbatim}
./scripts/run_all.sh
\end{verbatim}
Outputs are placed in \texttt{output/ae\_<timestamp>/}. Each figure or table can also be generated independently via a standalone script:

{\small
\begin{itemize}[nosep,leftmargin=*]
  \item \texttt{./scripts/run\_figure\_overall.sh} $\rightarrow$ Figure~8
  \item \texttt{./scripts/run\_figure\_nonsquare.sh} $\rightarrow$ Figure~9
  \item \texttt{./scripts/run\_figure\_mapping.sh} $\rightarrow$ Figure~10
  \item \texttt{./scripts/run\_figure\_breakdown.sh} $\rightarrow$ Figure~11
  \item \texttt{./scripts/run\_figure\_crossbar\_width.sh} $\rightarrow$ Figure~12(a)
  \item \texttt{./scripts/run\_figure\_window\_size.sh} $\rightarrow$ Figure~12(b)
  \item \texttt{./scripts/run\_k\_reordering.sh} $\rightarrow$ \textsection IV-C ablation result
\end{itemize}
}

\subsection*{Evaluation and Expected Results}

Because the simulator is fully deterministic, the CSV outputs should be bit-identical to the reference files shipped in \texttt{expected\_results/}. The generated plots faithfully reproduce Figures~8--12 of the paper.

In addition, the \texttt{hardware/} directory ships SystemVerilog RTL for every SegFold module together with synthesis reports (area, power, and timing) produced by Synopsys Design Compiler targeting the ASAP 7nm standard cell library. These correspond directly to the hardware cost analysis in the paper.

\subsection*{Experiment Customization}

Every experiment script supports \texttt{-{}-jobs N} for parallelism control, \texttt{-{}-config PATH} for alternative configurations, and \texttt{-{}-timeout SEC} for per-run time limits. A single matrix can be evaluated directly:
\begin{verbatim}
./csegfold/build/csegfold \
  --config configs/segfold.yaml \
  --mtx-file benchmarks/data/suitesparse/
    ca-GrQc/ca-GrQc.mtx
\end{verbatim}

\end{document}